\PassOptionsToPackage{pdfpagelabels}{hyperref}
\documentclass[fleqn,usenatbib]{mnras}

\usepackage{newtxtext,newtxmath}

\usepackage[T1]{fontenc}
\usepackage{ae,aecompl}

\usepackage{graphicx}	
\usepackage{amsmath,cuted,flushend}	
\usepackage{amssymb}	
\usepackage{bm}	
\usepackage{threeparttable, soul}
\usepackage{natbib}	

\def\tvi(#1,#2){\vrule height #1pt depth #2pt width 0pt}
\def\p{\partial}
\def\e{{\rm e}}
\def\d{{\rm d}}

\def\rso{r_{\rm s}}

\def\cs{c_{\rm s}}
\def\M{{\cal M}}

\def\eiwv{\e^{i\omega\int_R^r {\d r\over \upsilon}}}

\title[Acoustic waves in collapsing stars]{Acoustic wave generation in collapsing massive stars with convective shells}

\author[E. Abdikamalov and T. Foglizzo]{
Ernazar Abdikamalov$^{1}$\thanks{E-mail: ernazar.abdikamalov@nu.edu.kz} and
Thierry Foglizzo$^{2}$\thanks{E-mail: foglizzo@cea.fr}
\\
$^{1}$Department of Physics, Nazarbayev University, Nur-Sultan 010000, Kazakhstan\\
$^{2}$AIM, CEA, CNRS, Universit\'e Paris-Saclay, Universit\'e Paris Diderot, Sorbonne Paris Cit\'e, F-91191 Gif-sur-Yvette, France
}

\date{Accepted XXX. Received YYY; in original form ZZZ}

\pubyear{2019}

\begin{document}
\label{firstpage}
\pagerange{\pageref{firstpage}--\pageref{lastpage}}
\maketitle

\begin{abstract}
The convection that takes place in the innermost shells of massive stars plays an important role in the formation of core-collapse supernova explosions. Upon encountering the supernova shock, additional turbulence is generated, amplifying the explosion. In this work, we study how the convective perturbations evolve during the stellar collapse. Our main aim is to establish their physical properties right before they reach the supernova shock. To this end, we solve the linearized hydrodynamics equations perturbed on a stationary background flow. The latter is approximated by the spherical transonic Bondi accretion, while the convective perturbations are modeled as a combination of entropy and vorticity waves. We follow their evolution from large radii, where convective shells are initially located, down to small radii, where they are expected to encounter the accretion shock above the proto-neutron star. Considering typical vorticity perturbations with a Mach number $\sim 0.1$ and entropy perturbations with magnitude $\sim 0.05 k_\mathrm{b}/\mathrm{baryon}$, we find that the advection of these perturbations down to the shock generates acoustic waves with a relative amplitude $\delta p/\gamma p \lesssim 10\%$, in agreement with published numerical simulations. The velocity perturbations consist of contributions from acoustic and vorticity waves with values reaching $\sim 10\%$ of the sound speed ahead of the shock. The perturbation amplitudes decrease with increasing $\ell$ and initial radii of the convective shells.
\end{abstract}



\section{Introduction}
\label{sec:intro}

The strong convection that massive stars develop in their innermost nuclear-burning shells are expected to play an important role in their explosions \cite[e.g.,][]{couch:15b,mueller:17}. Following the collapse of the iron core, the convective perturbations descend from their initial position at $\gtrsim 1500 \,\mathrm{km}$ towards the center of the star. The supernova shock, launched at core bounce, encounters these perturbations at a radius of $\sim \! 150 \,\mathrm{km}$ within $\sim \! 200-300\,\mathrm{ms}$ after formation (or within $\sim \! 400-500\,\mathrm{ms}$ after the start of the iron core collapse) \citep[e.g.,][]{mueller:15,mueller:16b}. The interaction of the two amplifies the violent non-radial motion in the post-shock region, generating an additional pressure behind the shock and thus creating a more favorable condition for producing an explosion \citep{couch:13d,couch:15b,takahashi:16,mueller:17,nagakura:19}. The oxygen-burning and, to a lesser extent, the silicon-burning shells are expected to have a particularly strong impact on the explosion condition \citep{collins:18}. 

During their accelerated infall towards the shock, the convective perturbations undergo profound evolution, as revealed by multi-dimensional numerical simulations \citep{buras:06b,mueller:15,couch:15b,mueller:17} as well as semi-analytical \citep{takahashi:14} and analytical calculations \citep{kovalenko:98,lai:00}. The density of the collapsing shells increases as they descend towards the center. The infall velocity gradually increases, becoming supersonic in the inner part of the flow. The shrinking convective vortices spin up due to the conservation of angular momentum. In addition, the convective eddies have to constantly adjust to new pressure balance, a process that generates strong acoustic waves \citep[e.g.,][]{foglizzo:01}. When these perturbations arrive ahead of the supernova shock, their physical properties affect the way they interact with the shock \citep{abdikamalov:16,abdikamalov18,huete:18,huete:19,radice:18}.

The aim of our work is to shed some light on the physical properties of the convective perturbations right before they reach the supernova shock. We treat the convective perturbations as a combination of vorticity and entropy waves co-moving with the mean flow. We evolve the perturbations using an extension of the linear hydrodynamics formalism of \citet{foglizzo:01}. Our work improves on previous studies in a number of ways. We follow the evolution of the perturbations starting from their initial location at $\gtrsim 1.5\times 10^3\,\mathrm{km}$ down to regions with radii $\sim 150\,\mathrm{km}$ where they are expected to encounter the supernova shock. Thus, we go beyond the $r\rightarrow 0$ asymptotic limit used in the previous works \citep{kovalenko:98,lai:00}. In addition, the simplicity of our method allows us to obtain an additional insight into the physics of the process compared to three-dimensional numerical simulations \citep{couch:15b,mueller:17}. In particular, we establish the physical constituents of the perturbations -- the vorticity, entropy, and acoustic waves -- and calculate their properties. 

The paper is organized as following. We present the method in Section~\ref{sec:method}. The results are presented in Section~\ref{sec:results}. The conclusion is provided in Section~\ref{sec:conclusion}.

\section{Method}
\label{sec:method}

\begin{figure}
\begin{center}
 \includegraphics[angle=0,width=1\columnwidth,clip=false]{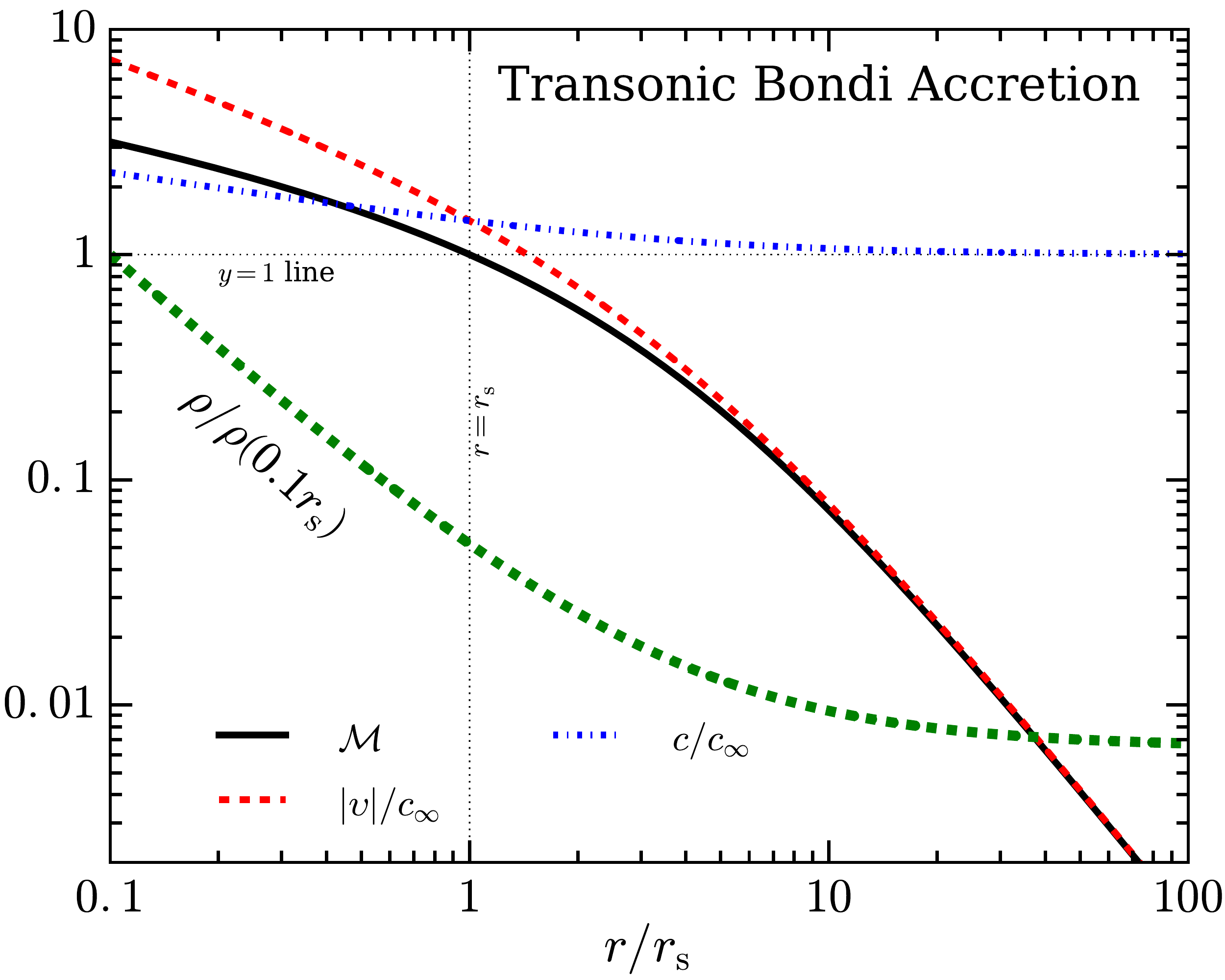}
  \caption{Transonic Bondi solution as a function of radius for $\gamma=4/3$. The thick black lines shows the Mach number of the flow, while the dashed thick red line shows the advection velocity in units of $c_\infty$. The sound speed is shown with dashed-dotted blue line. For reference, the thin vertical dotted line shows the location of the sonic point $r=\rso$, while the thin horizontal dotted line shows the ordinate $y=1$. 
  \label{fig:bondi}}
\end{center}
\end{figure}

\begin{figure*}
\begin{center}
 \includegraphics[angle=0,width=1.5\columnwidth,clip=false]{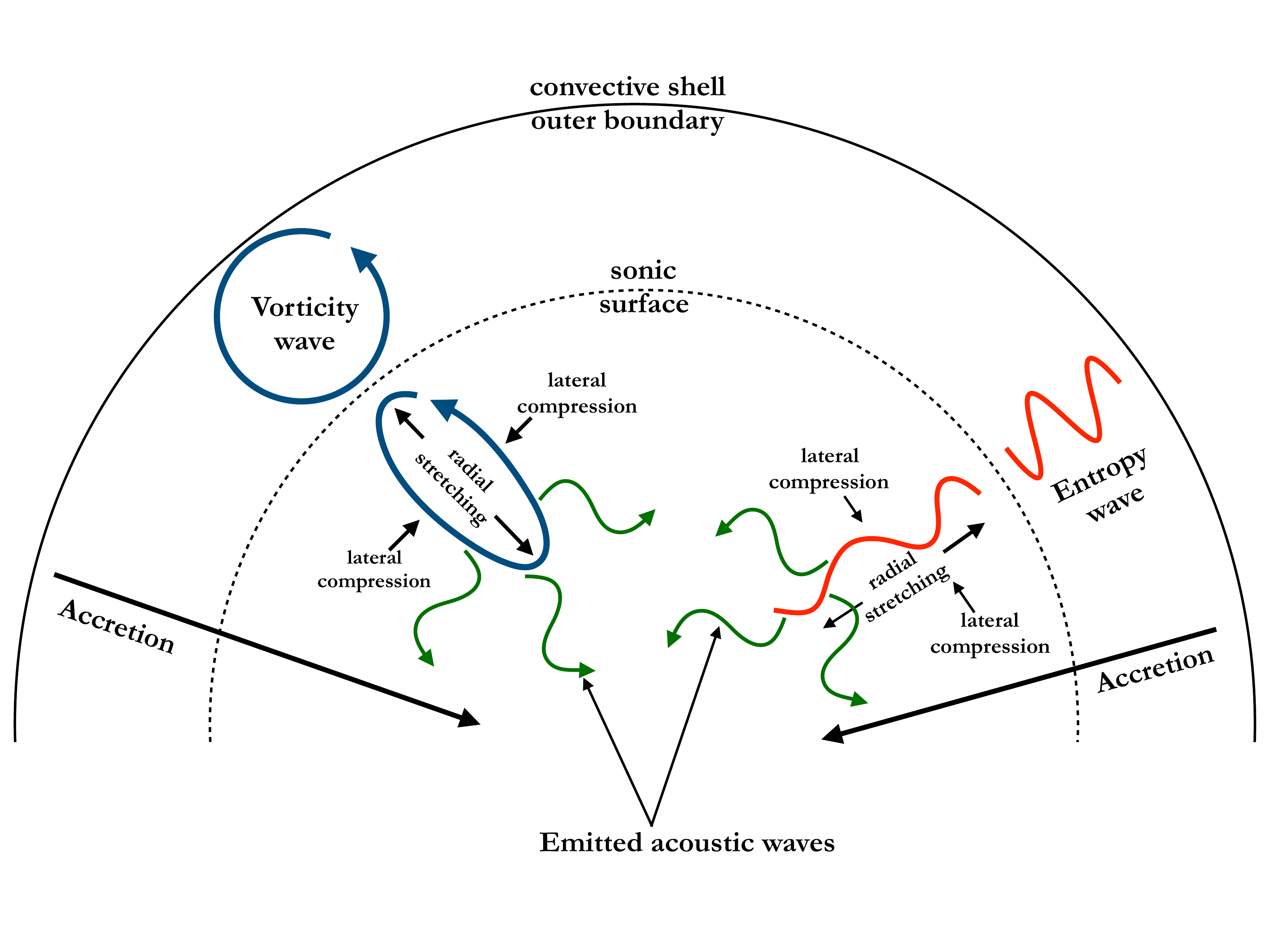}
  \caption{Approximate schematic depiction of horizontal vorticity and entropy waves in convective shells of a collapsing star. During the collapse, these perturbations are advected towards the center together with the flow. The contraction of the waves generates pressure perturbations that travel as acoustic waves. The contracting entropy waves generate additional vorticity via the baroclinic effect. At large radii, the infall velocity is small, but as the collapse progresses down to small radii, the infall speed accelerates (cf. Fig~\ref{fig:bondi}) and becomes supersonic. The sonic surface is shown with the dashed semi-circle. The entropy and vorticity perturbations are radially stretched by the acceleration. Note that both vorticity and entropy waves couple to pressure perturbations even while traveling in the subsonic region, but their amplitude is much smaller and hence it is not depicted here for the clarity of the illustration.  
  \label{fig:scheme}}
\end{center}
\end{figure*}

We solve the linearized hydrodynamics equations for advected convective perturbations on a stationary background flow. The stellar matter is modeled using an ideal gas equation of state with an adiabatic index $\gamma=4/3$. In order to check the sensitivity of our results to the value of $\gamma$, we perform additional calculations for $\gamma=1.5$ and we find qualitatively similar results. We assume that the background flow is given by the stationary spherical transonic Bondi solution \citep{bondi52}. The radial profiles of velocity, speed of sound, density, and Mach number are shown in Fig.~\ref{fig:bondi}. The mean flow speed increases with decreasing $r$. While the Bondi solution represents an approximation to the flow in realistic stars (e.g., it assumes uniform composition and neglects the time-dependence of the collapse), its simplicity allows us to obtain a unique and deep insight into the physics of the advection of the perturbations. The flow is subsonic (supersonic) above (below) the sonic radius $\rso$,
\begin{equation}
\rso = {5-3\gamma \over 4} r_\mathrm{B},
\label{rson}
\end{equation}
where $r_\mathrm{B}$ is the Bondi radius $GM/c_\infty^2$ and $c_\infty$ is the sound speed at infinity, which is a free parameter in our model. Due to the stationarity of the background flow, the mass of the accretor $M$ is assumed to be constant in our model. That said, all the results that we present in this work are independent of the values of $M$ and $c_\infty$. 
At the sonic point $\rso$, the flow velocity equals the local sound speed,
\begin{equation}
\cs=\left({2\over5-3\gamma}\right)^{1\over2} c_\infty.
\label{cson}
\end{equation}
For $\gamma=4/3$, the sound speed at the sonic point equals $\sqrt{2} c_\infty$. Details of the Bondi solution are described in the Appendix A of \citet{foglizzo:01}. 

We model convective perturbations as a combination of vorticity and entropy waves. Since the convection in nuclear-burning shells is subsonic \citep[e.g.,][]{kippenhahn:13}, the contribution of acoustic waves is considered negligible before collapse \citep{lighthill:52,lighthill:54,goldreich:90}. We also neglect internal gravity waves in our model. While g-modes are expected to play an important role in stellar evolution \citep[e.g.,][]{quataert:12,fuller:17} and may affect the final spin of the stellar core \citep{fuller:15spin}, their impact on the explosion condition of CCSNe are expected to be rather minor \citep{mueller:17}. 

We decompose the velocity field of hydrodynamic perturbations into vector spherical harmonics \citep[e.g.,][]{chatzopoulos:14,kovalenko:98}
\begin{eqnarray}
\label{eq:vdecompose1}
\delta \bm{\upsilon}(r,t,\theta,\phi) &=& \bigg\{ \delta \upsilon_r(r)  Y_{\ell m} \, \boldsymbol{\hat r} \\\nonumber \\\nonumber
&+& \delta \upsilon_\bot(r) L^{-1} \boldsymbol{\hat \nabla}_\bot Y_{\ell m} \\\nonumber \\\nonumber
&+& \delta \upsilon_\mathrm{rot}(r) L^{-1} \boldsymbol{\hat \nabla}_\bot Y_{\ell m} \times  \, {\bm{\hat r}} \bigg\} 
\mathrm{e}^{-\mathrm{i} \omega t},
\end{eqnarray}
where $\omega$ is the angular frequency, $\hat {\bm{r}}$, $\hat {\bm{\theta}}$, and $\hat {\bm{\phi}}$ are unit vectors and
\begin{equation}
\label{eq:nabla_t1}
\boldsymbol{\hat \nabla}_\bot = \hat {\bm \theta} \frac{\p}{\p \theta} + \hat {\bm \phi} \frac{1}{\sin\theta}. 
\end{equation}
The normalization factor $L = (\ell(\ell+1))^{1/2}$ is introduced to account for the asymptotic amplitude of the angular derivative of the spherical harmonic $Y_{\ell m}$. We show in Appendix \ref{sec:deltaK} that the horizontal component $\delta \upsilon_\mathrm{rot}(r)$ fully accounts for the vertical component of vorticity. In the linear approximation, $\delta \upsilon_\mathrm{rot}(r)$ decouples from the rest of the flow and scales as $\propto r^{-1}$, as dictated by the conservation of angular momentum. The radial and transverse components $\delta \upsilon_r(r)$ and $\delta \upsilon_\bot(r)$ are responsible for the horizontal components of the vorticity vector as well as acoustic waves.

For adiabatic flows, the entropy variations are conserved and "frozen into" the mean flow. The amplitude of vorticity perturbations ${\boldsymbol\delta \omega \equiv {\boldsymbol\nabla} \times {\boldsymbol\delta \upsilon}}$ is affected by advection and by entropy perturbations in such a way that the quantity $\delta K$ defined in \citet{foglizzo:01} is linearly conserved and acts as a source for the generation of sound waves (cf. Appendix \ref{sec:deltaK}):
\begin{equation}
\delta K\equiv r^2 {\boldsymbol \upsilon} \cdot ({\boldsymbol\nabla}\times {\boldsymbol\delta \omega})+L^2 c^2 {\delta S\over\gamma},\label{eq:defK}
\end{equation}
where $\delta S$ is the dimensionless entropy, the value of which equals the entropy per baryon in the units of Boltzmann constant $k_\mathrm{b}$, as shown in Appendix \ref{sec:entropy}. 
Following \cite{foglizzo:01}, we model both perturbations as sinusoidal waves with frequency $\omega$ that are advected with the mean flow. Thus, the incoming perturbations are characterized by only four quantities: the amplitudes $|\delta K|$ and $|\delta S|$ associated to the frequency $\omega$ and the angular wavenumber $\ell$. 

We formulate the linear hydrodynamics equations in a compact form using the function $\delta \tilde f$, which is related to the perturbations of the Bernoulli constant of the flow (cf. Appendix~\ref{sec:hydro_lin}):
\begin{equation}
\label{canonic0}
{\p^2 \delta \tilde f\over\p X^2}+W \delta \tilde f = A \, {\delta S_R} + B \, {\delta K_R}
\end{equation}
where the variable $X$ is related to $r$ via Eq.~(\ref{eq:x}), while the functions $W$, $A$, and $B$ are related to the properties of the background flow as well as the frequency $\omega$ and wavenumber $\ell$ of the perturbations (cf. eqs.~\ref{canonic}-\ref{defW}). The quantities ${\delta S_R}$ and ${\delta K_R}$ are set by the amplitudes of entropy and horizontal vorticity perturbations at the radius $R$. Thus, the solution of the equation is linearly proportional to the amplitude of the source terms ${\delta S_R}$ and ${\delta K_R}$. The homogeneous part of Eq.~(\ref{canonic0}) describes freely propagating acoustic waves. The general solution of Eq.~(\ref{canonic0}) is obtained in Appendices \ref{sec:hydro_lin}-\ref{sec:acoustic} using Green functions and the regularity condition at the sonic point. A second-order Frobenius expansion is necessary to smoothly connect the solutions in the subsonic and supersonic regions. Far from the accretor, the identification of ingoing and outgoing waves using the WKB approximation allows us to define the outer boundary condition as the absence of incoming acoustic waves from infinity. We set the outer boundary at $40\rso$, which is sufficiently far away for the WKB approximation to be valid (cf. Appendix~\ref{sec:homogeneous_solution}). An additional condition follows from the requirement of the regularity of the solution at the sonic point. The numerical solutions of the homogeneous equation are obtained using an implicit Runge-Kutte method. 

The angular wavenumber of the dominant mode is largely determined by the size of the shell relative to its radius \citep{chandra:61,foglizzo:06}:
\begin{equation} 
\ell \sim \frac{\pi}{2} \frac{ r_{+} + r_{-} }{ r_{+} - r_{-} },
\label{eq:l}
\end{equation} 
where $r_{+}$ and $ r_{-} $ are the outer and inner boundaries of the convective shell. Modes with $\ell$ ranging from 1 to $\sim 100$ have been observed in numerical simulations \citep{collins:18}, but the impact of large-$\ell$ modes on the explosion condition of CCSNe are expected to be rather limited \citep{mueller:16,kazeroni19}. Assuming that the dominant mode with wavenumber $\ell$ spans the entire radial extent of the convective zone, the radial size $\Delta R = r_{+} - r_{-}$ is related to $\ell$ using Eq.~(\ref{eq:l}):
\begin{equation} 
\Delta R  = \frac{\pi}{\ell} \frac{r_{+} + r_{-}}{2} = \frac{\pi}{\ell} R_\mathrm{shell},
\end{equation} 
where $R_\mathrm{shell}$ is interpreted as the radius of the convective vortices before collapse. The value of $R_\mathrm{shell}$ is different for different stellar models \cite[e.g.,][]{collins:18}. For this reason, we treat $R_\mathrm{shell}$ as a free parameter. In our model, this correponds to a radius at which the vortex has a circular shape. For the Bondi flow, this condition can be expresssed as $R_\mathrm{shell}=L\upsilon(R_\mathrm{shell})/\omega$, where $\upsilon(R_\mathrm{shell})$ is the advection velocity at $r=R_\mathrm{shell}$. In our study, we consider $4$ values of $R_\mathrm{shell}$ ranging from $\rso$ to $4\rso$, keeping in mind that $\rso \sim 2,000-3,000\,\mathrm{km}$ within $\sim 100-300\,\mathrm{ms}$ after bounce, a time span within each CCSN shock is expected to encounter the convective perturbations. As we will see below, these values of $R_\mathrm{shell}$ cover the parameter space of the perturbations that are likely to have the strongest impact on the explosion condition of CCSNe.

Numerical simulations predict convective Mach numbers $ \lesssim 0.1$ in the innermost shells \citep{mueller:16,collins:18,yadav19,yoshida19}, while the associated entropy fluctuations are $ \lesssim 0.05 \, k_\mathrm{b}/\mathrm{nucleon}$ \citep[e.g.,][]{meakin:07}. In our calculations, we normalize entropy perturbations to $0.05 k_\mathrm{b}/\mathrm{baryon}$. For horizontal vorticity perturbations, we choose the normalization factor, i.e. $|\delta K|$, to yield a convective Mach number of $0.1$ at the pre-collapse locations of the convective vortices. Since the shell convection in massive stars is driven by buoyancy, the vertical vorticity is expected to be smaller than the horizontal component. This is supported by the study of \citet{chatzopoulos:14}, who performed the decomposition of convective velocities in massive stars into vector spherical harmonics. At peak $\ell$, their parameter $\beta$, which measures the horizontal circulation, is smaller by a factor of a few than their measure of vertical circulation $\gamma$ (cf. Fig.~13 of \citet{chatzopoulos:14}). In our study, we conservatively assume that the Mach number of the horizontal circulation is $\delta \upsilon_\mathrm{rot} = 10^{-2}$, i.e., an order of magnitude smaller than that of the vertical circulation. We point out that, while the precise values of $\delta \upsilon_\mathrm{rot}$, $\delta K$, and $\delta S$ are likely to depend on the properties of specific stellar models \citep[e.g.,][]{collins:18}, our formalism is linear with respect to $\delta \upsilon_\mathrm{rot}$, $\delta K$, and $\delta S$. Hence, our results can simply be scaled linearly to other values of these parameters. This allows us to capture any value of the convective Mach numbers and entropy fluctuations. 

In summary, the values of $|\delta K|$ and $|\delta S|$ as well as $\ell$ and the initial radius $R_\mathrm{shell}$ of convective vortices fully specify the physical properties of the convective perturbations which couple to the pressure field in our model.

\section{Results}
\label{sec:results} 

\subsection{Qualitative picture}
\label{sec:qualitative}

The production of pressure perturbations from the advection of horizontal vorticity perturbations can be understood by considering a vorticity perturbation $\delta \omega$ with a characteristic size $\delta r$ in a collapsing star. As it moves together with the converging mean flow, this perturbation distorts the iso-density surfaces of the flow and induces a density change \citep{foglizzo:01,mueller:15}. This density change is associated with pressure perturbation $\delta p /\gamma p \sim \delta \rho / \rho$. To an order of magnitude, 
\begin{equation}
\frac{\delta p}{\gamma p} \sim \frac{\delta \rho}{\rho} \sim \frac{\partial \ln \rho}{\partial \ln r} \frac{\delta r}{r} 
\end{equation}
where $\rho$ is the mean density and $p$ is the mean density of the background flow. The displacement $\delta r $ is related to the radial velocity perturbations via $\delta r \sim 2\pi\delta \upsilon_r / \omega$, where $\omega$ is the angular frequency of the perturbation.  The radial velocity perturbation $\delta \upsilon_r$ is related to the perturbed vorticity $\delta \omega$ via $\delta \omega \sim i m \delta \upsilon_r / r$, where $m$ is the angular order of the perturbation. Combining these, we obtain 
\begin{equation}
\label{eq:deltp_acoustic_qual}
\frac{\delta p}{\gamma p} \sim \frac{\partial \ln \rho}{\partial \ln r} \frac{2\pi \delta \omega}{i m \omega}.
\end{equation}
The pressure perturbation $\delta p / \gamma p $ is thus expected to be largest for small $m$, i.e., for large-scale perturbations \citep{foglizzo:09,mueller:15}. In the limit of a uniform flow (${\partial \ln \rho} / {\partial \ln r}=0$), the advection of vorticity perturbations does not emit acoustic waves as expected \citep{kovasznay:53}. Note that the emission of sound by advected horizontal vorticity can also be explained using the shallow water analogy \citep{foglizzo:15}. The vertical component of vorticity does not couple to the pressure field.

The production of pressure perturbations from the advection of entropy perturbations can be understood by considering a fluid element of mass $m$ with a perturbed entropy $\delta s$. The expansion of a gas element under an adiabatic change of pressure depends on its entropy. The corresponding change of volume induces the emission of acoustic waves. When the fluid element is advected from a region with mean specific enthalpy $h_1$ to another region with mean specific enthalpy $h_2$, the energy of the emitted acoustic waves is deduced from energy conservation \citep{foglizzo:00}
\begin{equation}
\label{eq:e21sound}
\delta E = \left(h_2-h_1 \right) \delta m,
\end{equation}
where
\begin{equation}
\label{eq:deltam}
\delta m = m \frac{\delta \rho}{\rho} = m \frac{\delta s}{\gamma c_\mathrm{v}}
\end{equation}
is the the variation of the mass $m$ of the fluid element with same volume and perturbed entropy $\delta s$ and $c_\mathrm{v}$ is the specific heat at constant volume. From this, we can obtain the total specific energy of emitted acoustic waves
\citep{foglizzo:00}:
\begin{equation}
\label{eq:e21sound2}
\delta {\cal E} \sim \left(h_2-h_1 \right) \frac{\delta s}{\gamma c_\mathrm{v}}.
\end{equation}
Thus, the energy of sound waves is proportional to the entropy change $\delta s$ and to the variation $h_2-h_1$, of the enthalpy. No acoustic waves are emitted if the flow is uniform ($h_2=h_1$). A schematic depiction of the process is presented in Fig.~\ref{fig:scheme}.

In addition, if the entropy perturbations have a transverse structure, the surfaces of constant pressure do not coincide with those of constant density. The net pressure force on a fluid element does not pass through its center of mass. This baroclinic effect creates a net torque on the fluid element, generating additional vorticity \citep[e.g.,][]{ThorneBlandford17} as illustrated by Eqs.~(\ref{wtet}, \ref{wphi}) in Appendix~\ref{sec:deltaK}. 

\subsection{Evolution of vorticity}
\label{sec:vorticity}

\begin{figure*}
\begin{center}
\includegraphics[angle=0,width=\columnwidth,clip=false]{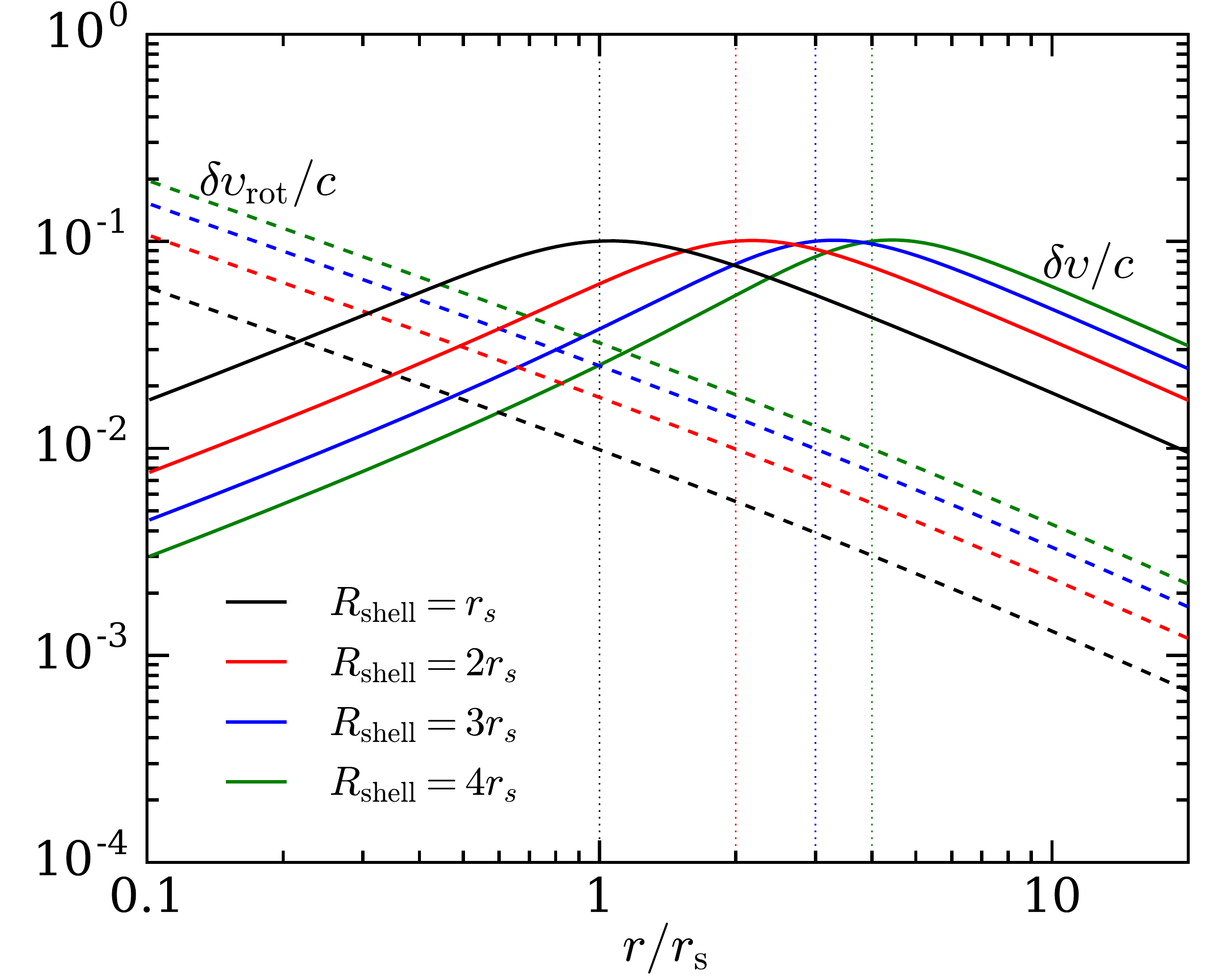}
\includegraphics[angle=0,width=\columnwidth,clip=false]{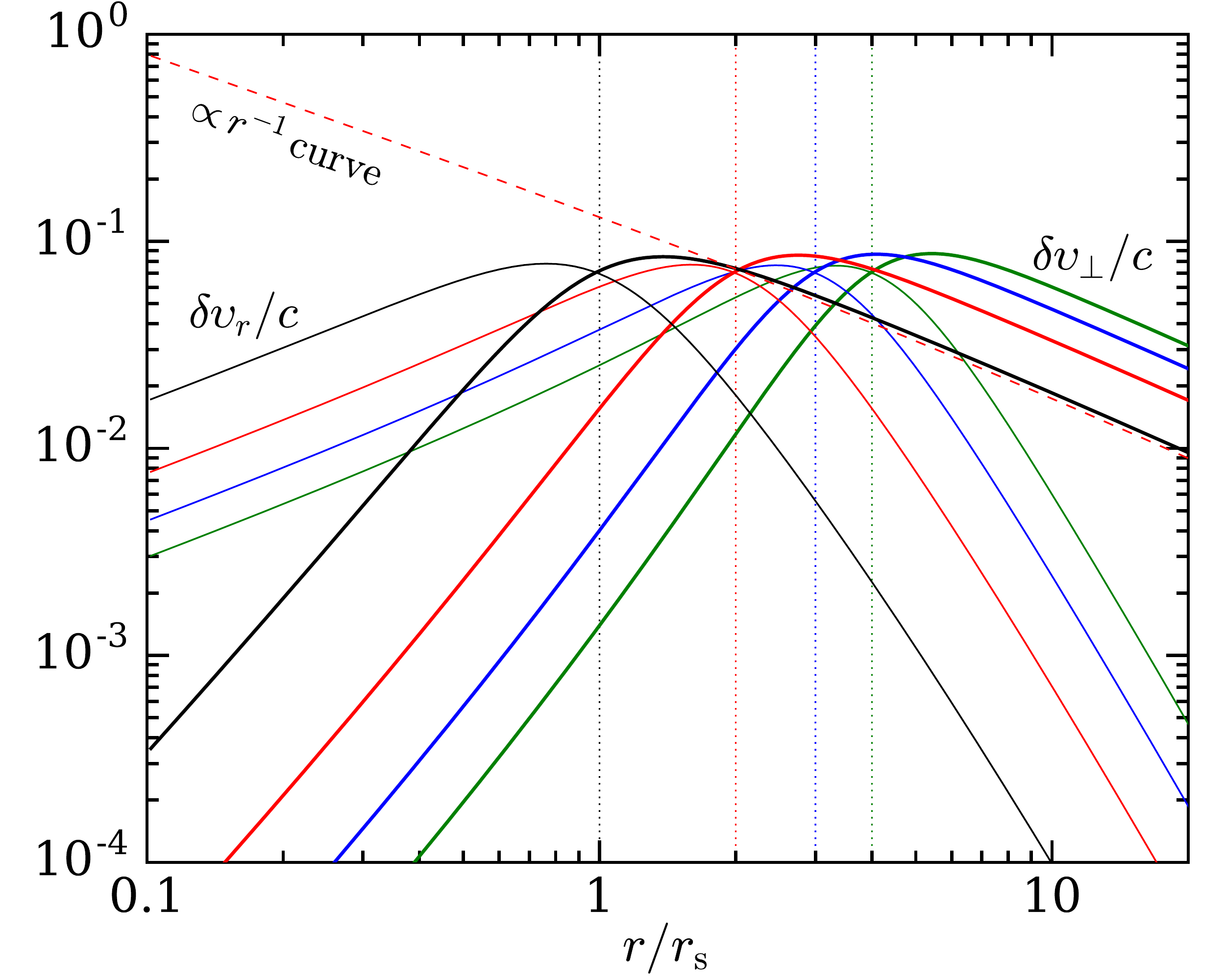}
  \caption{{\bf Left panel:} Mach number $\delta\upsilon/c$ of the advected horizontal vorticity waves as a function of radial coordinate $r$ for different values of the initial radii $R_\mathrm{shell}$ of the convective vortices. $\delta\upsilon$ is defined as $(\delta\upsilon_r^2+\delta\upsilon_\bot^2)^{0.5}$. {\bf Right panel:} Mach number of the transverse (think lines) and radial (thin lines) components of the velocity field of the advected horizontal vorticity waves as a function of $r$ for different values of $R_\mathrm{shell}$. The velocities are deduced from Eqs.~(\ref{eq:vr_vor}) and (\ref{eq:vt_vor}). The normalization factor for horizontal vorticity waves is chosen in such a way as to yield convective Mach number of $0.1$ at the pre-collapse location of the convective vortices $R_\mathrm{shell}$. The component $\delta \upsilon_\mathrm{rot}/c$, responsible for the vertical vorticity, is assumed to be $10^{-2}$ at $r=R_\mathrm{shell}$.
   \label{fig:dv_vorticity}}
\end{center}
\end{figure*}

\begin{figure}
\begin{center}
 \includegraphics[angle=0,width=\columnwidth,clip=false]{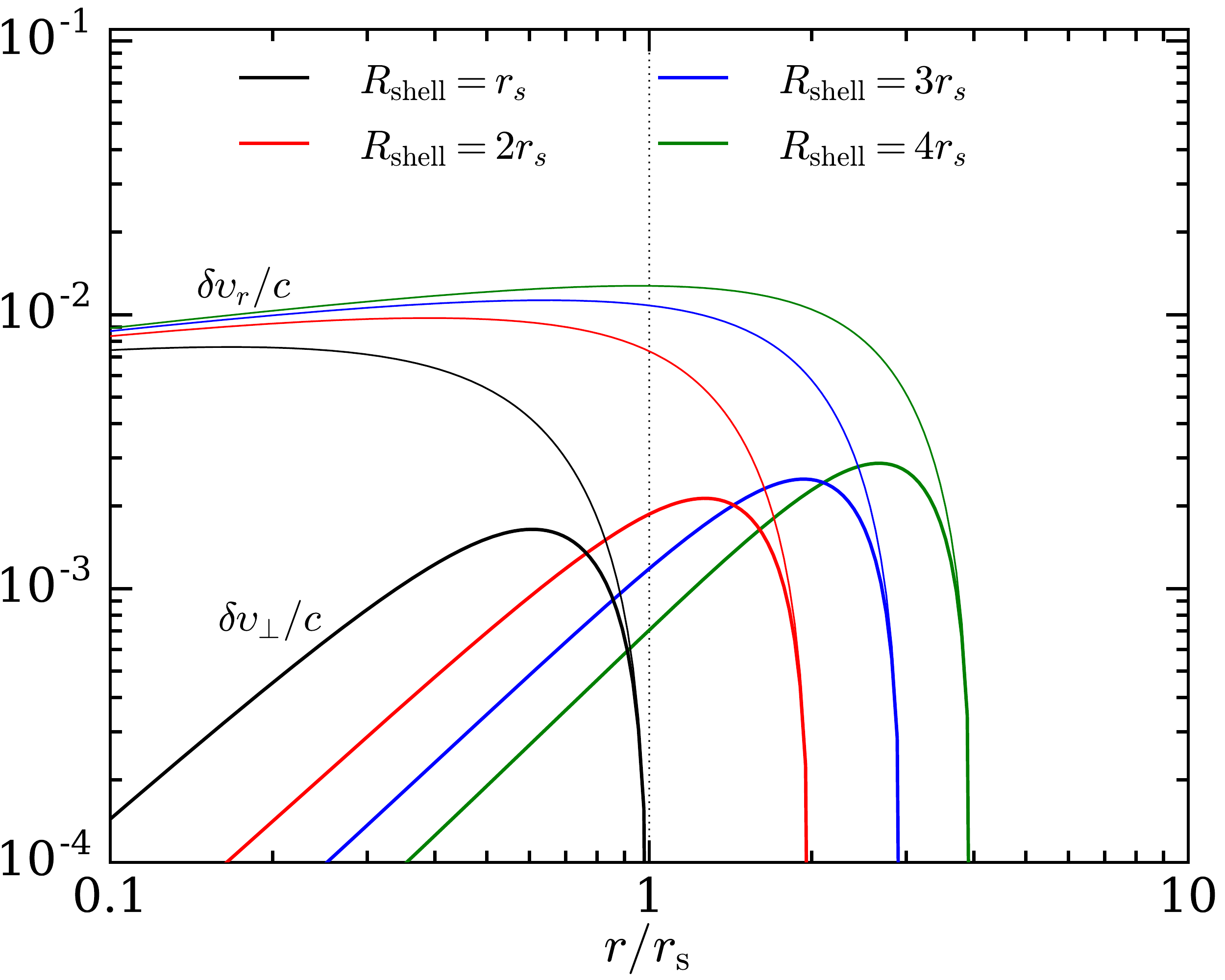}
  \caption{Mach number $\delta\upsilon/c$ of the transverse (thick lines) and radial components (thin lines) of the velocity field of horizontal vorticity perturbations generated by advected entropy fluctuations with $\delta S=0.05$ for different values of $R_\mathrm{shell}$. The velocities are deduced from Eqs.~(\ref{eq:vr_ent}) and (\ref{eq:vt_ent}). \label{fig:dv_entropy}}
\end{center}
\end{figure}

We now discuss how vorticity perturbations evolve during their advection towards the center. This includes not only the vorticity perturbations originating in convective shells, but also the vorticity generated by the advected entropy perturbations due to the baroclinic effect. After establishing the behavior of the vorticity perturbations, we will discuss the acoustic waves emitted by the advected horizontal vorticity and entropy perturbations (Section~\ref{sec:acoustic_waves}). 

As mentioned above, the velocity perturbation associated with the vertical component of vorticity evolves as $\propto r^{-1}$ \citep{kovalenko:98}. It is linearly decoupled from the rest of the perturbations and it does not depend on $\ell$. Depending on the initial radius, $\delta \upsilon_\mathrm{rot}$ may amplify of by a factor of $\gtrsim 10$ when they reach $r \sim 0.1 \rso$, which is approximately the radius where the stalled CCSN shock is expected to encounter these perturbations. The dashed lines on the left panel of Fig.~\ref{fig:dv_vorticity} shows $\delta \upsilon_\mathrm{rot}/c$ as a function of $r$ for different values of the initial radius $R_\mathrm{shell}$. The Mach number of horizontal motions associated to the vertical vorticity scales as $\delta \upsilon_\mathrm{rot}/c \sim 0.06 (R_\mathrm{shell}/\rso)^{0.9}$. For most values of $R_\mathrm{shell}$, $\delta \upsilon_\mathrm{rot}$ becomes $\sim 10^{-1}$ in the inner regions of the flow.   

The evolution of the horizontal vorticity is drastically different from that of the vertical component. The total velocity $\delta \upsilon$ associated to the horizontal vorticity, defined as $(\delta\upsilon_r^2+\delta\upsilon_\bot^2)^{0.5}$ and shown on the left panel of Fig.~\ref{fig:dv_vorticity}, decreases with radius starting from $r=R_\mathrm{shell}$. This decrease is caused by the stretching of vortex sheets in the radial direction by the accelerated mean flow. As the perturbation advects towards the center, its innermost point travels a longer distance than its outermost point. For this reason, the radial stretching increases more as the perturbation advects further down to smaller radii. The circulation of the vortex lines, defined as integral of velocity over a closed curve, 
\begin{equation}
\Gamma=\oint {\bf \upsilon} d {\bf s},
\end{equation}
is a conserved quantity when the entropy is uniform \citep[e.g.,][]{landau:59}. As the length of the closed curve increases due to the stretching of vortex sheets, the velocity along this curve has to decrease as observed in our calculations. Note that this effect is less pronounced for modes with small $R_\mathrm{shell}$, which we can see from the fact that velocities are larger for smaller $R_\mathrm{shell}$ at $r\lesssim \rso$ (cf. left panel of Fig.~\ref{fig:dv_vorticity}). This is not surprising as the modes with small $R_\mathrm{shell}$ have smaller radial sizes and thus are less stretched by the flow in the radial direction. Due to the radial stretching, the radial component of velocity becomes larger than the tangential component, as we can see on the left panel of Fig.~\ref{fig:dv_vorticity}. On the other hand, at large radii, the vortices are squeezed in the radial direction, and the tangential velocities dominate. Note that the velocities of the advected horizontal vorticity waves, once normalized to yield a Mach number of $0.1$ at $r=R_\mathrm{shell}$, do not depend on $\ell$.

An asymptotic analysis reveals that $\delta \upsilon_r \propto r^{1/2}$ and $\delta \upsilon_\bot \propto r^{2}$ in the limit $r \rightarrow 0$ (cf. Appendix~\ref{sec:decomposition}). Thus, the horizontal vorticity perturbations are expected to have a small velocity in this limit, while the vertical vorticity is expected to dominate due to the $\delta \upsilon_\mathrm{rot} \propto 1/r$ scaling. This result is in disagreement with \cite{kovalenko:98}, who find that $\delta \upsilon_r / c \propto r^{(3-3\gamma)/4}$ and $\delta \upsilon_\bot / c \propto r^{(3\gamma-7)/4}$ in the same limit, which results in $\delta \upsilon_r \propto r^{-1/2}$ and $\delta \upsilon_\bot \propto r^{-1}$ for $\gamma=4/3$. Their scaling appears to be valid for acoustic waves emitted by vorticity waves, not for the vorticity waves themselves (I. Kovalenko, private communication). This conclusion is supported by the fact that a similar scaling was obtained for acoustic waves for $r\rightarrow 0$ by \cite{lai:00}. 

The advected entropy waves generate horizontal vorticity due to the baroclinic effect, as mentioned above in Section~\ref{sec:qualitative}. Figure~\ref{fig:dv_entropy} shows the radial profile of $\delta \upsilon_\bot/c$ and $\delta \upsilon_r/c$ associated to the horizontal vorticity for different values of $R_\mathrm{shell}$. As with the advected vorticity waves, these velocities do no depend on $\ell$. Since the vortices are generated in radially-accelerated flow, the radial component dominates the transverse component, especially at small radii. An asymptotic analysis reveals that the tangential velocity decreases as $\delta \upsilon_\bot \propto r^{3/2}$ while the radial component approaches a constant value, $\delta \upsilon_r \sim \mathrm{10^{-2}}$ in the limit $r \rightarrow 0$ (cf. Appendix~\ref{sec:decomposition}). Thus, unlike the horizontal vorticity waves coming from convective shells, the vorticity generated by advected entropy waves has non-zero radial velocity even at $r\rightarrow0$. This is due to the fact that advected entropy waves continue to produce vorticity, through the baroclinic effect, even in the limit of small $r$. 

\subsection{Acoustic perturbations}
\label{sec:acoustic_waves}

\begin{figure*}
\begin{center}
 \includegraphics[angle=0,width=1\columnwidth,clip=false]{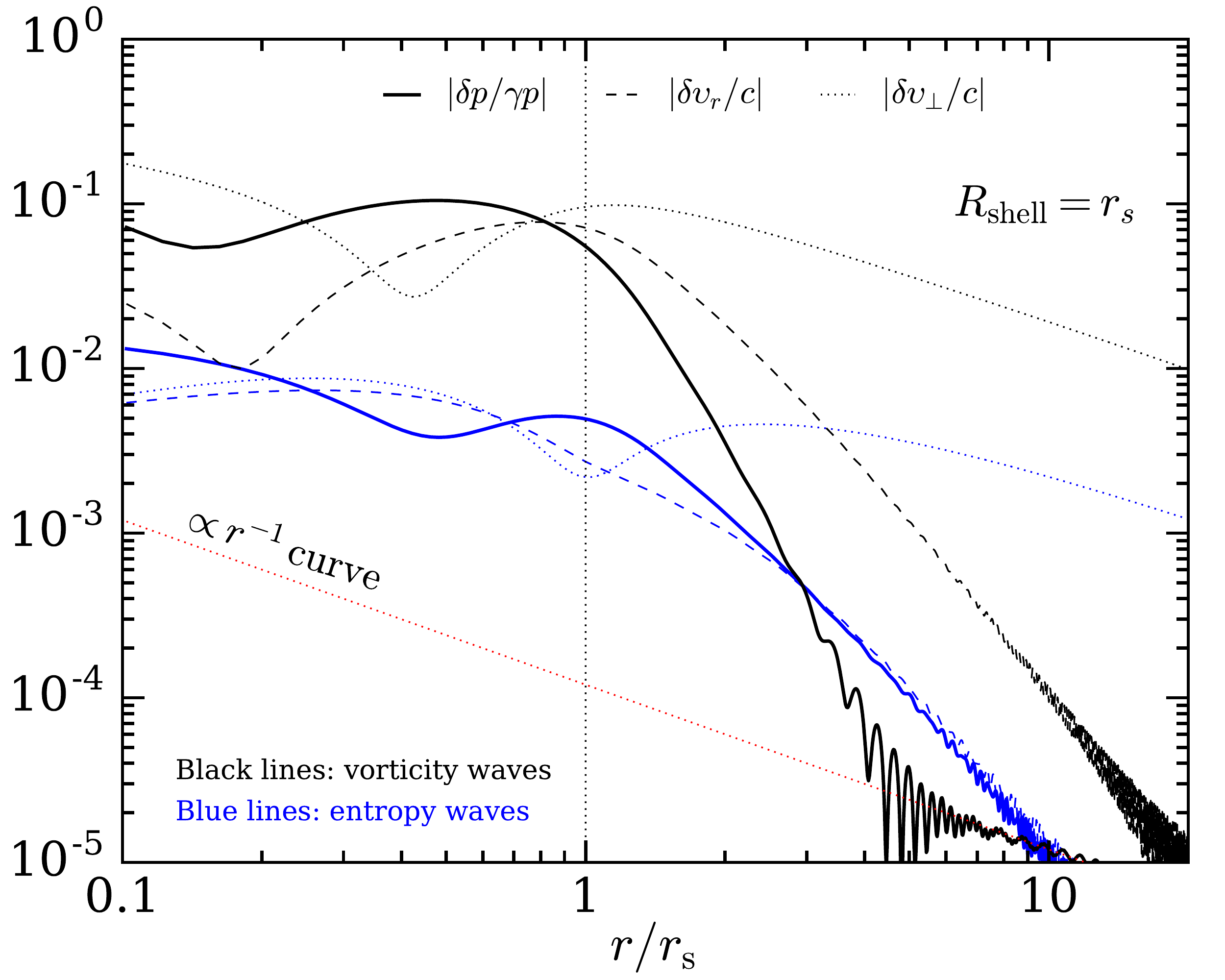}
 \includegraphics[angle=0,width=1\columnwidth,clip=false]{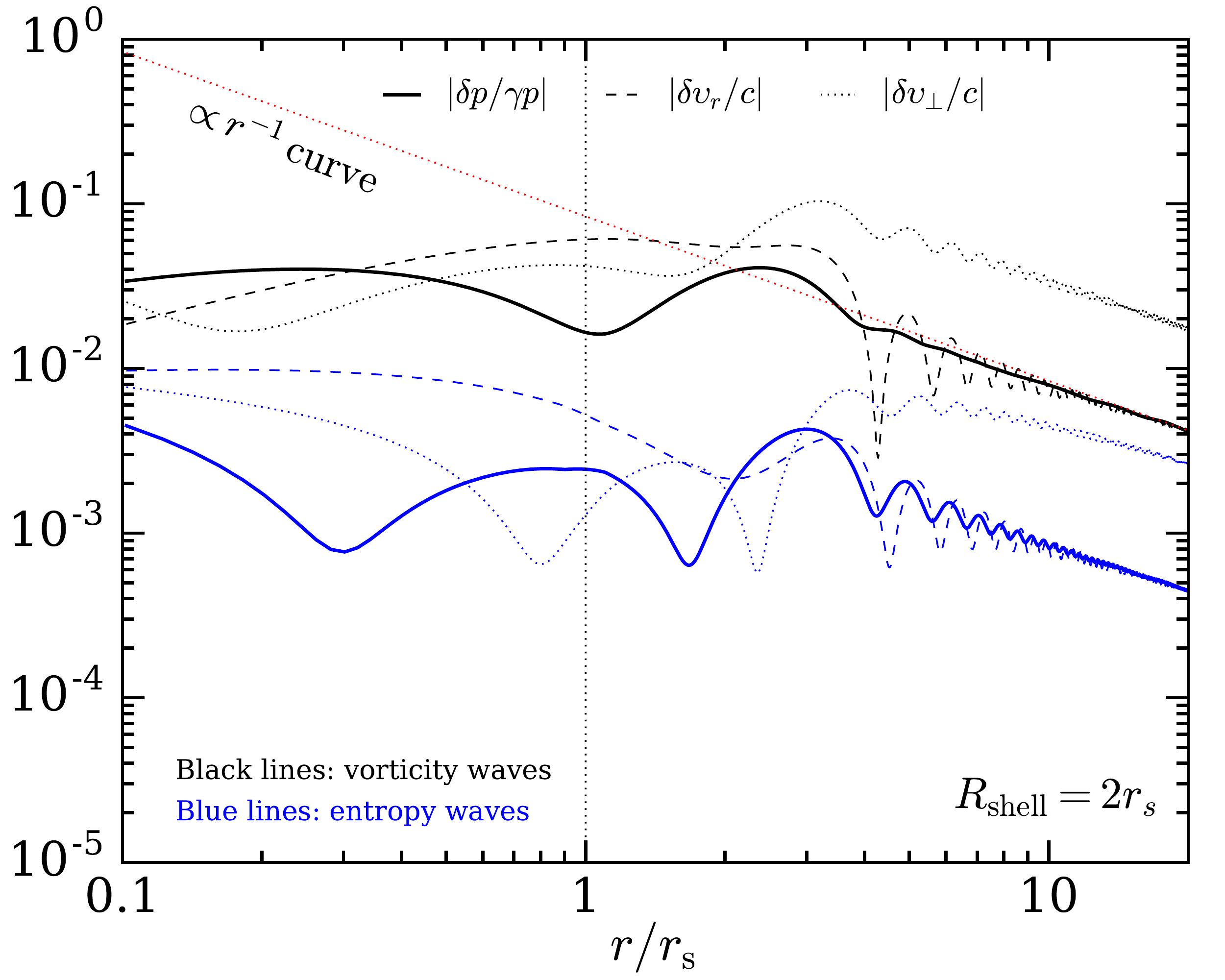}
  \caption{Pressure perturbations for incoming horizontal vorticity (thick black lines) and entropy (thick blue lines) waves with $\ell=2$ and $R_\mathrm{shell}=\rso$ (left panel) and $R_\mathrm{shell}=2\rso$ (right panel), deduced from Eq.~(\ref{defgp}). The dashed (dotted) lines show the amplitude of the radial velocity fluctuations $|\delta \upsilon_r / c|$ (tangential velocity fluctuations $|\delta \upsilon_\bot / c|$) for incoming horizontal vorticity and entropy waves. The vertical dashed line shows the location of the sonic point, while the dotted red line shows the $\propto r^{-1}$ slope for reference. 
  \label{fig:p_fudicial}}
\end{center}
\end{figure*}

\begin{figure*}
\begin{center}
 \includegraphics[angle=0,width=\columnwidth,clip=false]{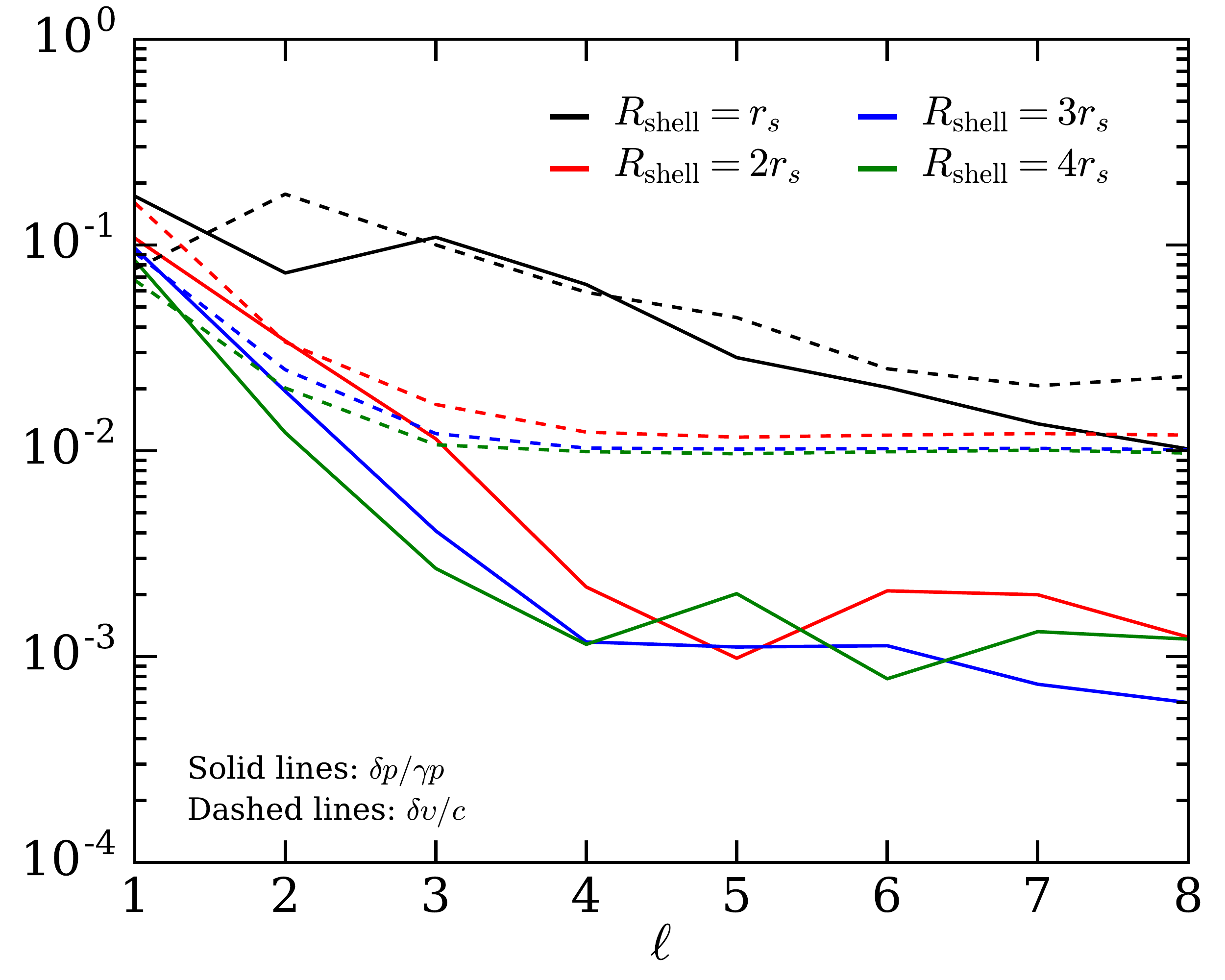}
 \includegraphics[angle=0,width=\columnwidth,clip=false]{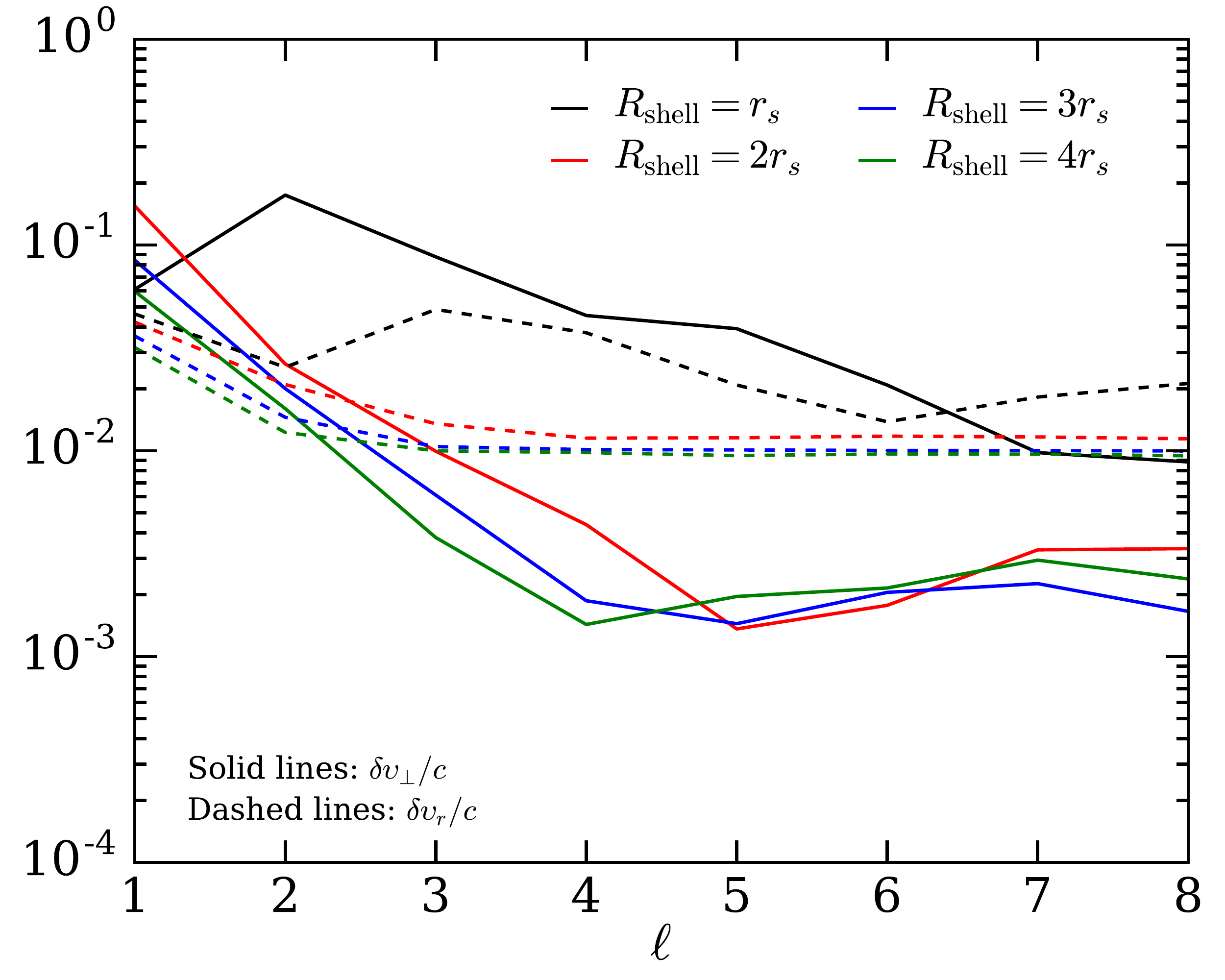}
  \caption{{\bf Left panel:} Pressure (solid lines) and velocity (dashed lines) perturbations at $r=0.1\rso$ generated by advected horizontal vorticity and horizontal vorticity as a function of angular wavenumber $\ell$ for different values of $R_\mathrm{shell}$. The pressure perturbations are deduced from Eq.~(\ref{defgp}), while velocity perturbations are dediced from Eqs.~(\ref{deffvt}) and (\ref{deffvr}). {\bf Right panel:} Transverse and radial velocity perturbations $\delta\upsilon_\bot/c$ (solid lines) and $\delta\upsilon_r/c$ (dashed lines) generated by advected horizontal vorticity waves as a function of $\ell$ for different values of $R_\mathrm{shell}$ at $r=0.1\rso$. 
  \label{fig:p_all}}
\end{center}
\end{figure*}

As the entropy and horizontal vorticity perturbations are advected towards the center, they generate acoustic waves due to the loss of pressure equilibrium with their surrounding. While in Section~\ref{sec:qualitative} we derived basic qualitative estimates, below we provide more quantitative results obtained by numerical integration of the differential system of perturbed equations (Appendix~\ref{sec:acoustic}).

Figure~\ref{fig:p_fudicial} shows the radial profiles of $|\delta p / \gamma p|$ generated by advected horizontal vorticity (black line) and entropy waves (blue line) with $\ell=2$ for two different value of $R_\mathrm{shell}$: $\rso$ (left panel) and $2\rso$ (right panel). Due to its smaller size, the model with $R_\mathrm{shell}=\rso$ undergoes less radial stretching than the model with $R_\mathrm{shell}=2\rso$. As a result, the former model reaches stronger pressure perturbations of $\sim 0.1 $ in the inner regions of the flow, while the latter remains below $\sim 4\times 10^{-2}$. For both models, the contributions of the advected entropy waves (blue lines) is about an order of magnitude smaller than that of advected horizontal vorticity waves. We find that this is common for perturbations with low $\ell=1$ and $\ell=2$, while for larger $\ell$ the contributions of both perturbations become comparable. The pressure perturbations experience little growth from $R_\mathrm{shell}$ to $0.1\rso$. This is different from the power-law dependence on radius obtained by \citet{kovalenko:98} and \citet{lai:00} in the $r\rightarrow 0$ limit. Instead, it resembles a wave-like pattern observed in \citet{takahashi:14} (cf. their Fig. 5). We hypothesize that this behavior is caused by the interference between ingoing and outgoing acoustic waves, forming a pattern similar to standing waves. However, we cannot verify this rigorously because the decomposition into out-going and in-going acoustic waves is not possible for non-uniform flow. In principle, this can be done using the WKB approximation (cf. Appendix~\ref{sec:decomposition}), but this approximation is accurate only in the outer region where the characteristic scale of the flow becomes larger than the size of the perturbations. 

At large radii ($r>R_\mathrm{shell}$), both models exhibit out-going acoustic waves due to the refraction of in-going radiation \citep{foglizzo:01}. The outgoing acoustic waves exhibit the $\propto r^{-1}$ scaling shown by the dotted red line, which is a simple consequence of the conservation of energy. The perturbation with $R_\mathrm{shell}=2\rso$ exhibits significantly stronger out-going radiation than the model with $R_\mathrm{shell}=\rso$. This is an expected behavior as waves with larger wavelength undergo stronger refraction \citet{foglizzo:01}.

It is interesting to contrast the behavior of the pressure perturbations with that of velocity perturbations. The dashed and dotted lines in Fig.~\ref{fig:p_fudicial} show $\delta \upsilon_r / c$ and $\delta \upsilon_\bot / c$. In the supersonic region ($r < \rso$), both quantities are comparable to the value of $\delta p/ \gamma p$ as expected for sound waves \citep[e.g.,][]{landau:59}. This suggests that the velocity field at small radius is mostly due to acoustic waves. At large radius ($r \gtrsim \rso$), the situation depends on the perturbation parameters. For the perturbation with $R_\mathrm{shell}=2\rso$, as discussed above, there is significant amount of out-going acoustic waves. In this case, we have $\delta \upsilon_r /c \sim \delta p /\gamma p$, as expected for acoustic waves. However, the transverse component is larger than the radial component by about an order of magnitude because, at large radii, the vertical vortices become squeezed in the radial direction and thus develop strong non-radial velocities. For perturbation with $R_\mathrm{shell}=\rso$, both $\delta \upsilon_r / c$ and $\delta \upsilon_\bot / c$ become significantly larger than $\delta p / \gamma p$. The reason for this behavior is that the velocity field at large radius is dominated by the contribution of vorticity waves only, while the contribution of acoustic waves is negligible. This weak advective-acoustic coupling at large radius is a consequence of the uniform character of the flow within distances comparable to the size of the perturbations. As in the case of pressure perturbations, the contribution of the advected entropy waves to $\delta \upsilon_r / c$ and $\delta \upsilon_\bot / c$, shown with blue dashed and dotted lines in Fig.~\ref{fig:p_fudicial}, is a factor $\sim 10$ smaller than the contribution of advected vorticity perturbations. 

Next we analyze the behavior of $\delta p / \gamma p$ at $0.1\rso$ generated the advected entropy and vertical vorticity perturbations\footnote{We superpose the pressure contributions of entropy $\delta p_e$ and vorticity waves $\delta p_k$ as $\delta p = (\delta p_e^2 + \delta p_k^2)^{0.5}$.}, which is shown on the left panel of Fig.~\ref{fig:p_all} for different values of $\ell$ and $R_\mathrm{shell}$. For small values of $\ell=1$ and $\ell=2$ or for $r=R_\mathrm{shell}$, we find that $|\delta p/\gamma p| \sim 0.1$, in agreement with the results of 3D numerical simulations \citep{mueller:17}. The pressure perturbations decreases with increasing $\ell$, in agreement with the qualitative estimate (\ref{eq:deltp_acoustic_qual}). It also decreases with increasing $R_\mathrm{shell}$, becoming $\lesssim 10^{-3}$ for $R_\mathrm{shell} = 4\rso$ and $\ell \gtrsim 4$. This is caused by the fact that the large-$R_\mathrm{shell}$ waves have larger size and thus are more prone to radial stretching due to the acceleration of the flow. This results in weaker velocity and pressure perturbations in the inner regions of the flow. The velocity perturbations $\delta \upsilon/c$ are comparable to $\delta p/\gamma p$ for $\ell \le 3$, as shown on the left panel of Fig.~\ref{fig:p_all}. For larger $\ell$, velocity perturbations $\delta \upsilon/c$ are much larger than pressure variations $\delta p / \gamma p$. This is because, for larger $\ell$, the generation of acoustic waves is not as efficient, so the velocity field is dominated by the contribution of the advected horizontal vorticity waves. The radial and tangential components of the velocity perturbations, shown on the left panel of Fig.~\ref{fig:p_all}, are comparable to each other for $\ell \lesssim 3$, but for larger $\ell$, the radial component tends to dominate for most values of $R_\mathrm{shell}$, which is expected for radially stretched vorticity waves. For reference, we provide the values of the amplitudes of the pressure and velocity perturbations at $r=0.1\rso$ generated by advected entropy and horizontal vorticity waves in Table~\ref{tab:models} for all of our perturbation parameters.

\begin{figure*}
\begin{center}
\includegraphics[angle=0,width=0.65\columnwidth, clip=false]{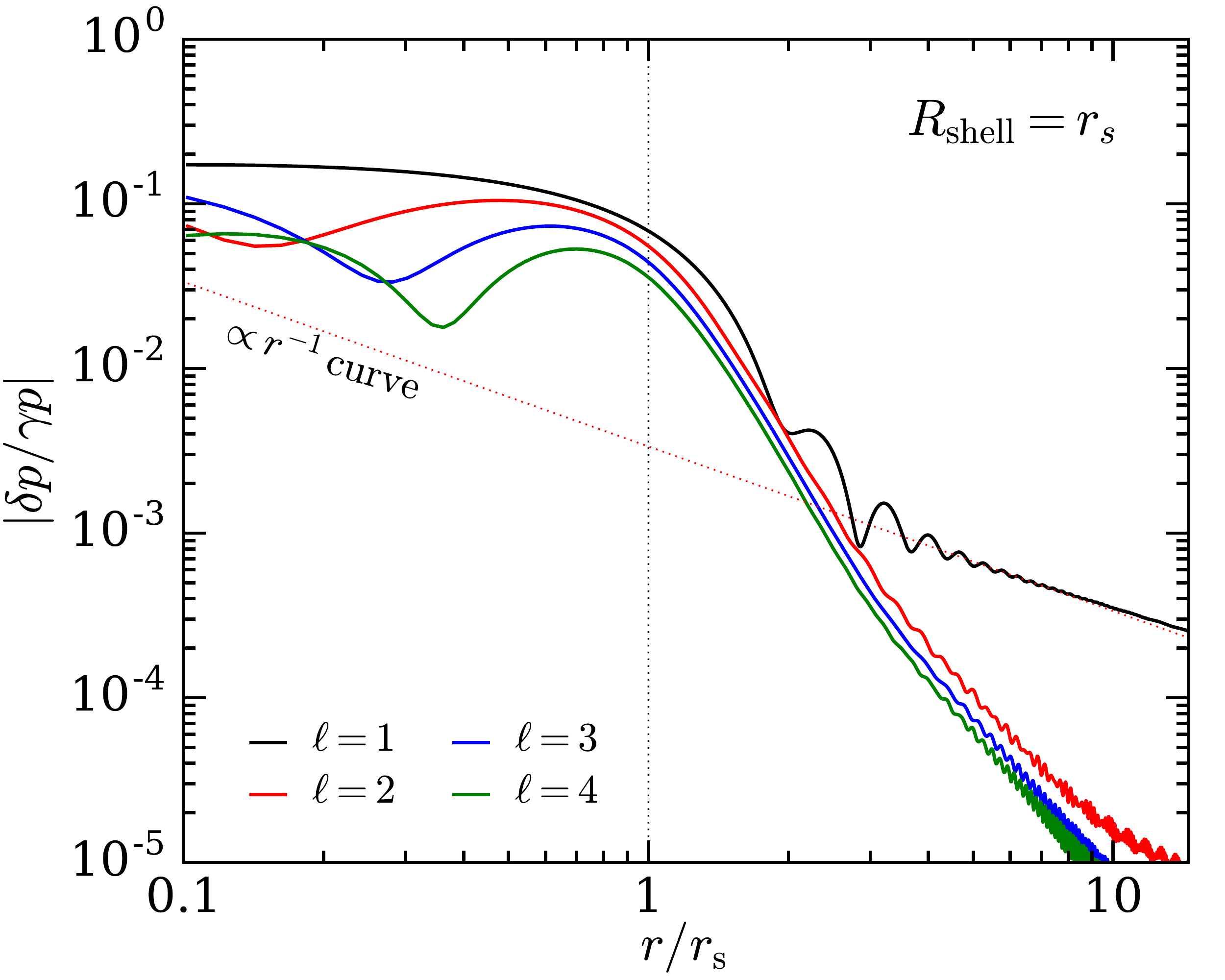}
\includegraphics[angle=0,width=0.65\columnwidth, clip=false]{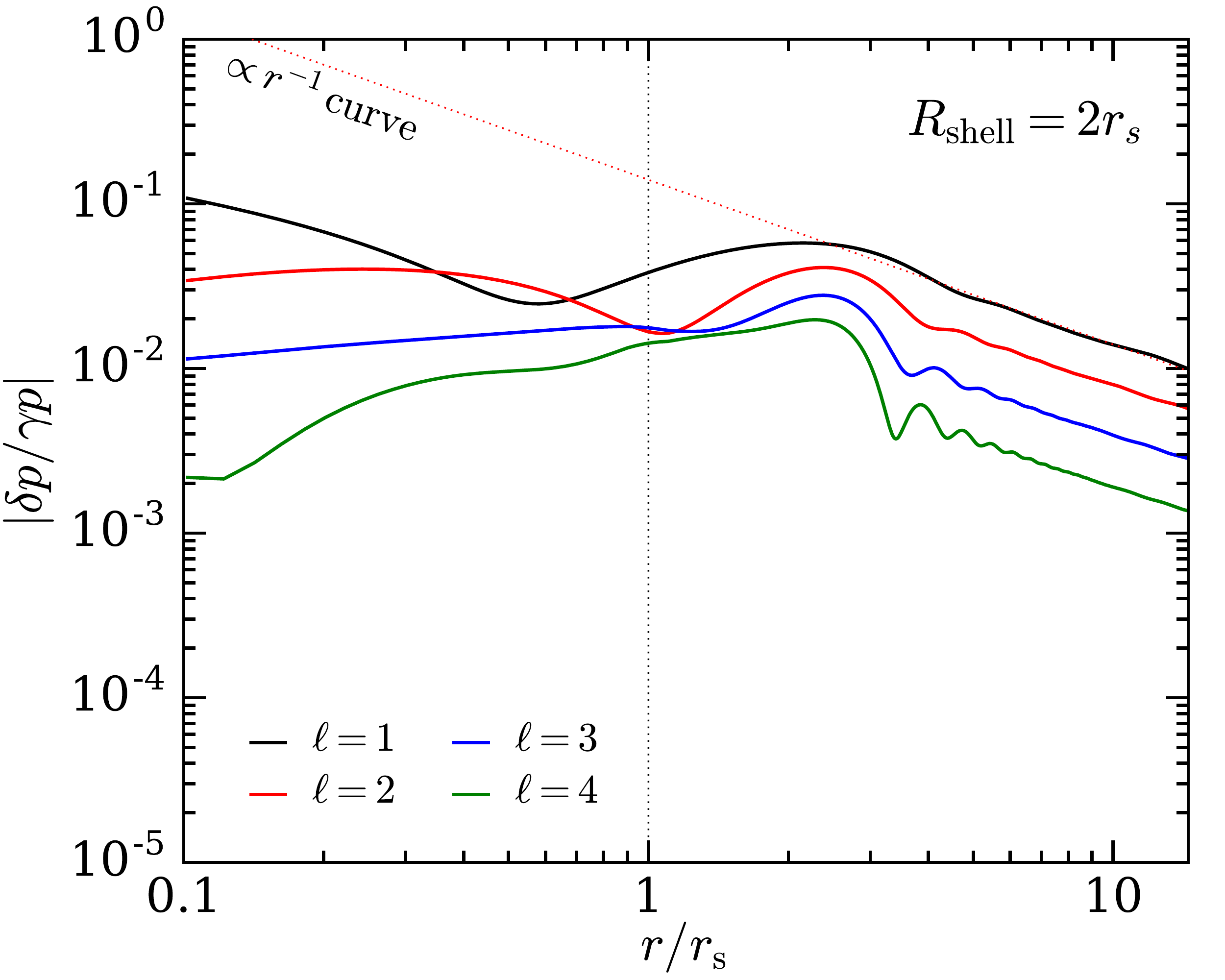}
\includegraphics[angle=0,width=0.65\columnwidth, clip=false]{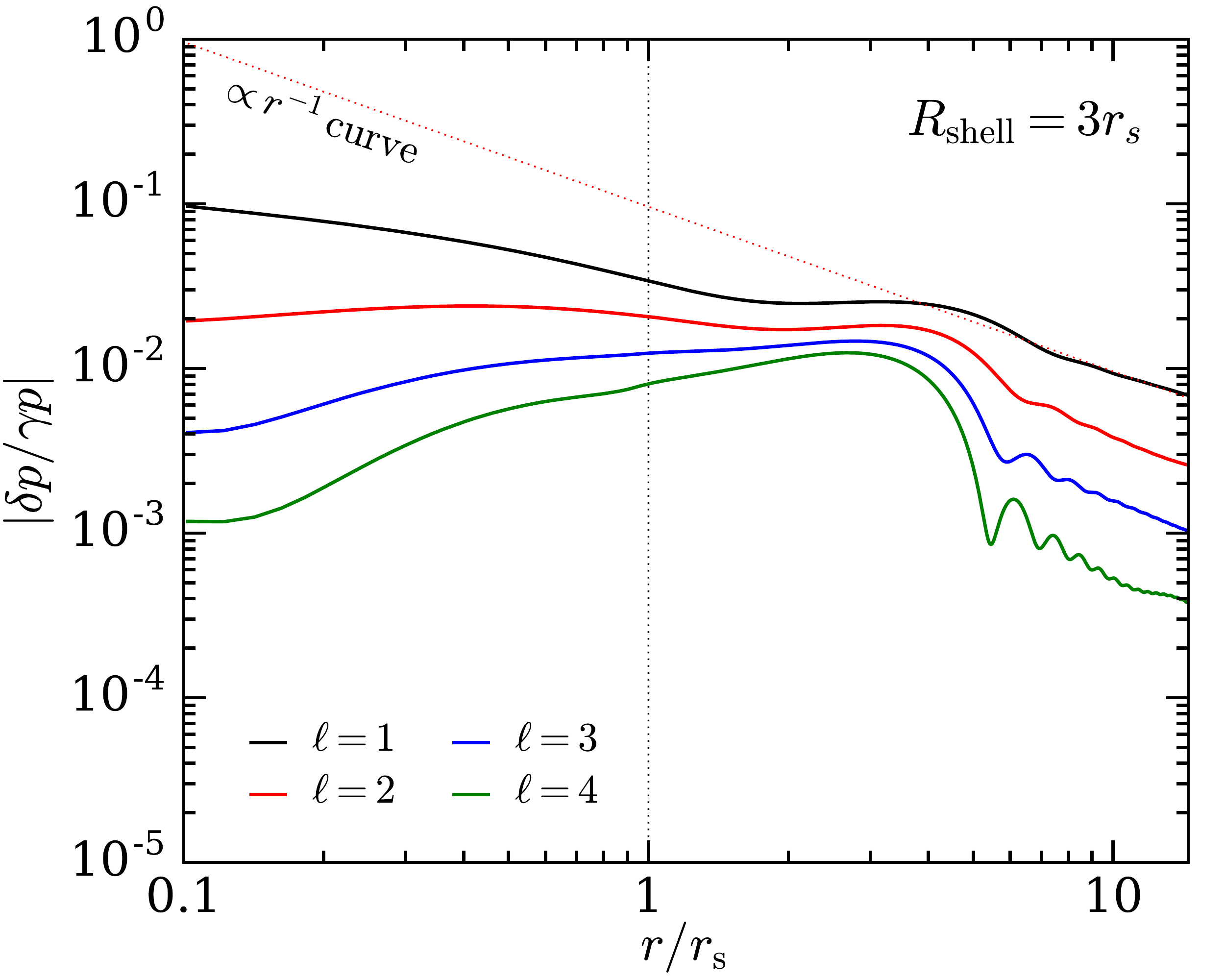}
\includegraphics[angle=0,width=0.65\columnwidth, clip=false]{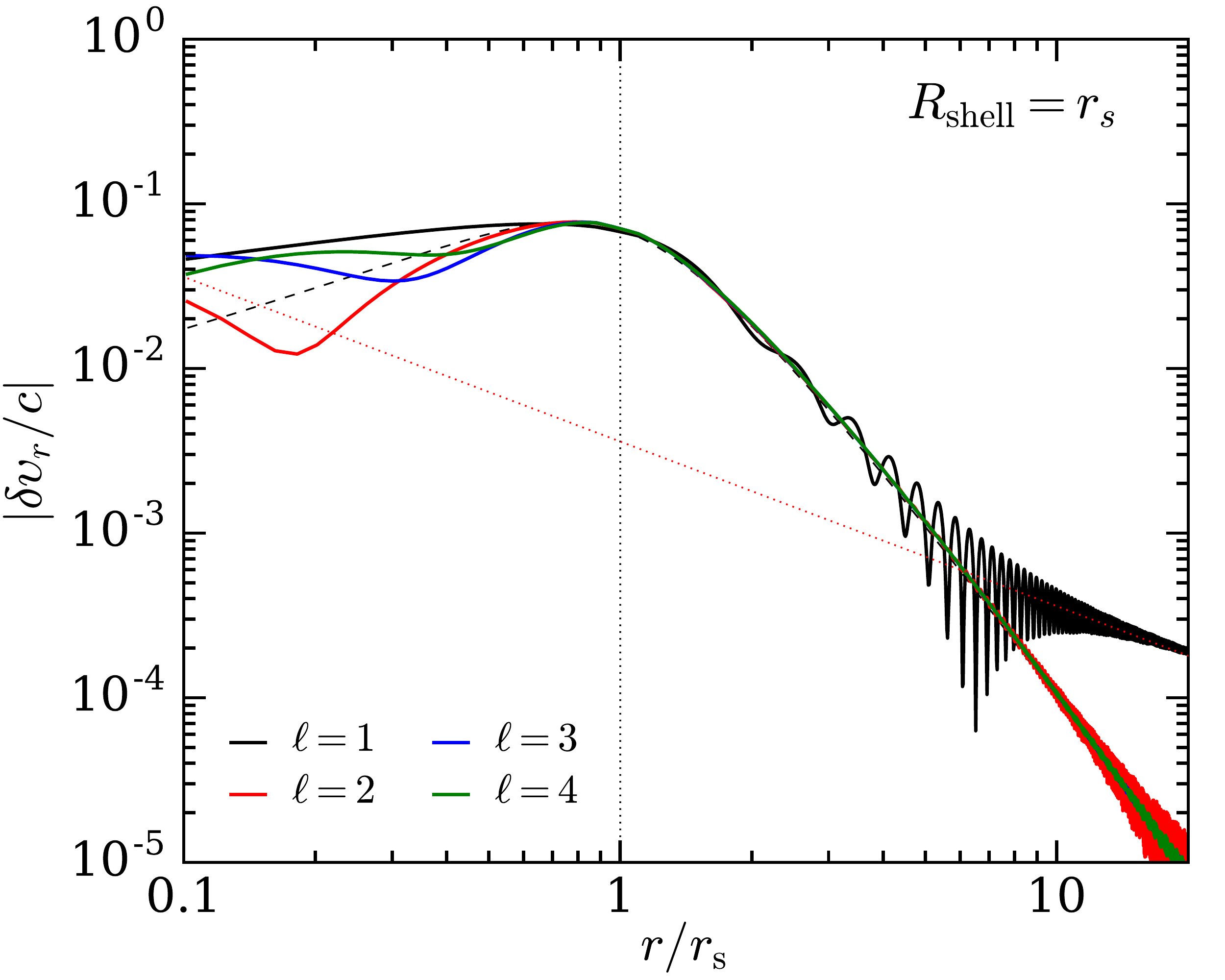}
\includegraphics[angle=0,width=0.65\columnwidth, clip=false]{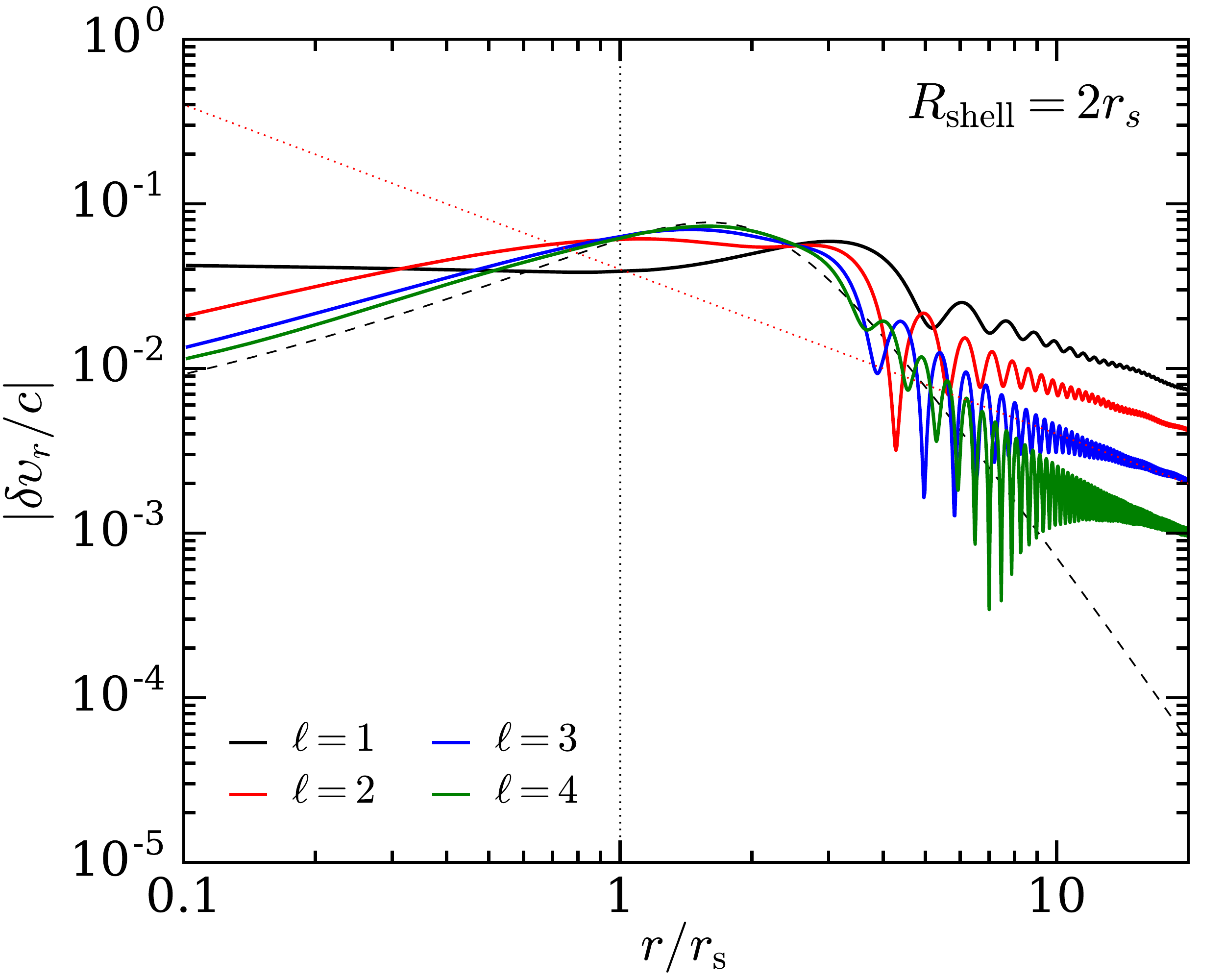}
\includegraphics[angle=0,width=0.65\columnwidth, clip=false]{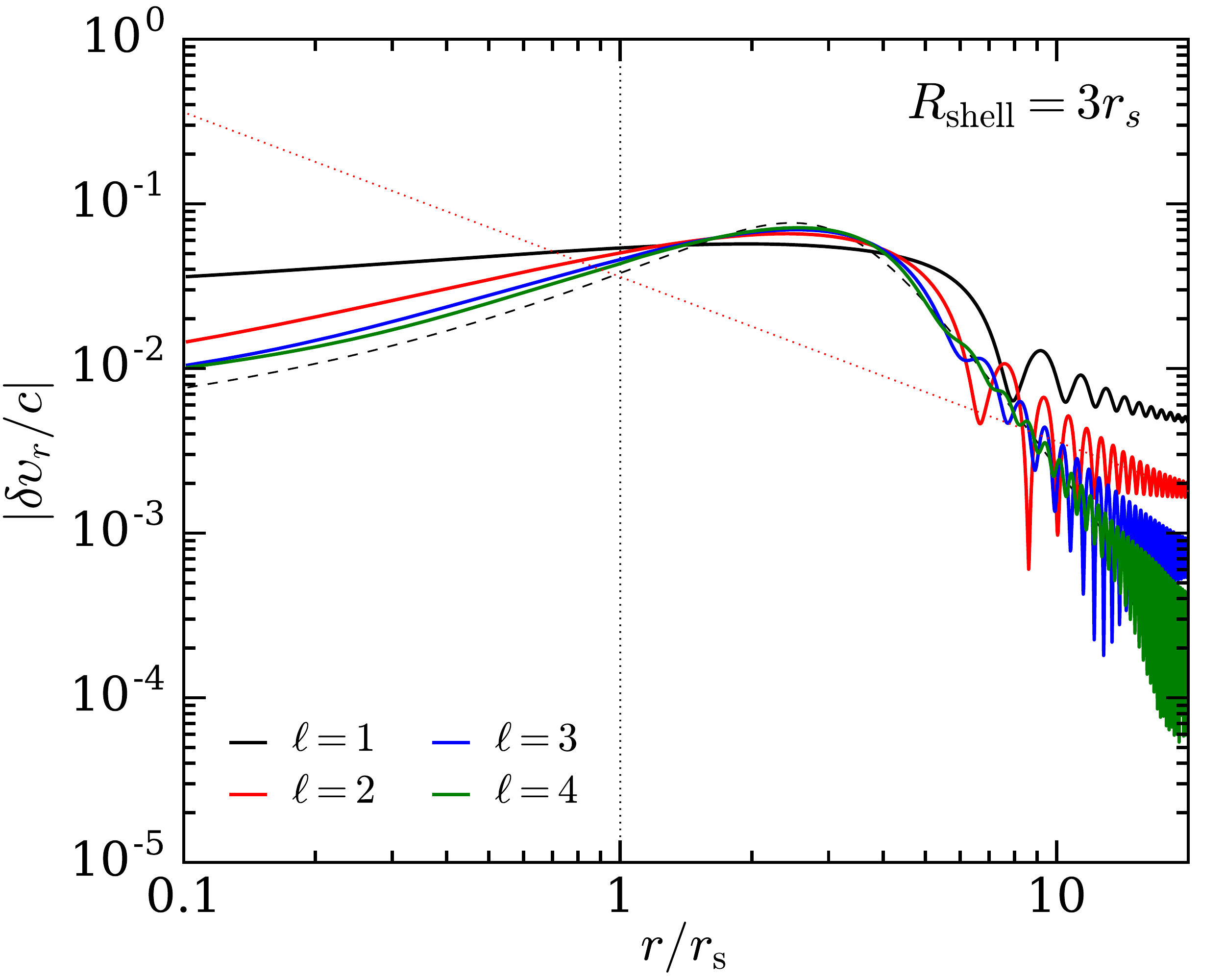}
\includegraphics[angle=0,width=0.65\columnwidth, clip=false]{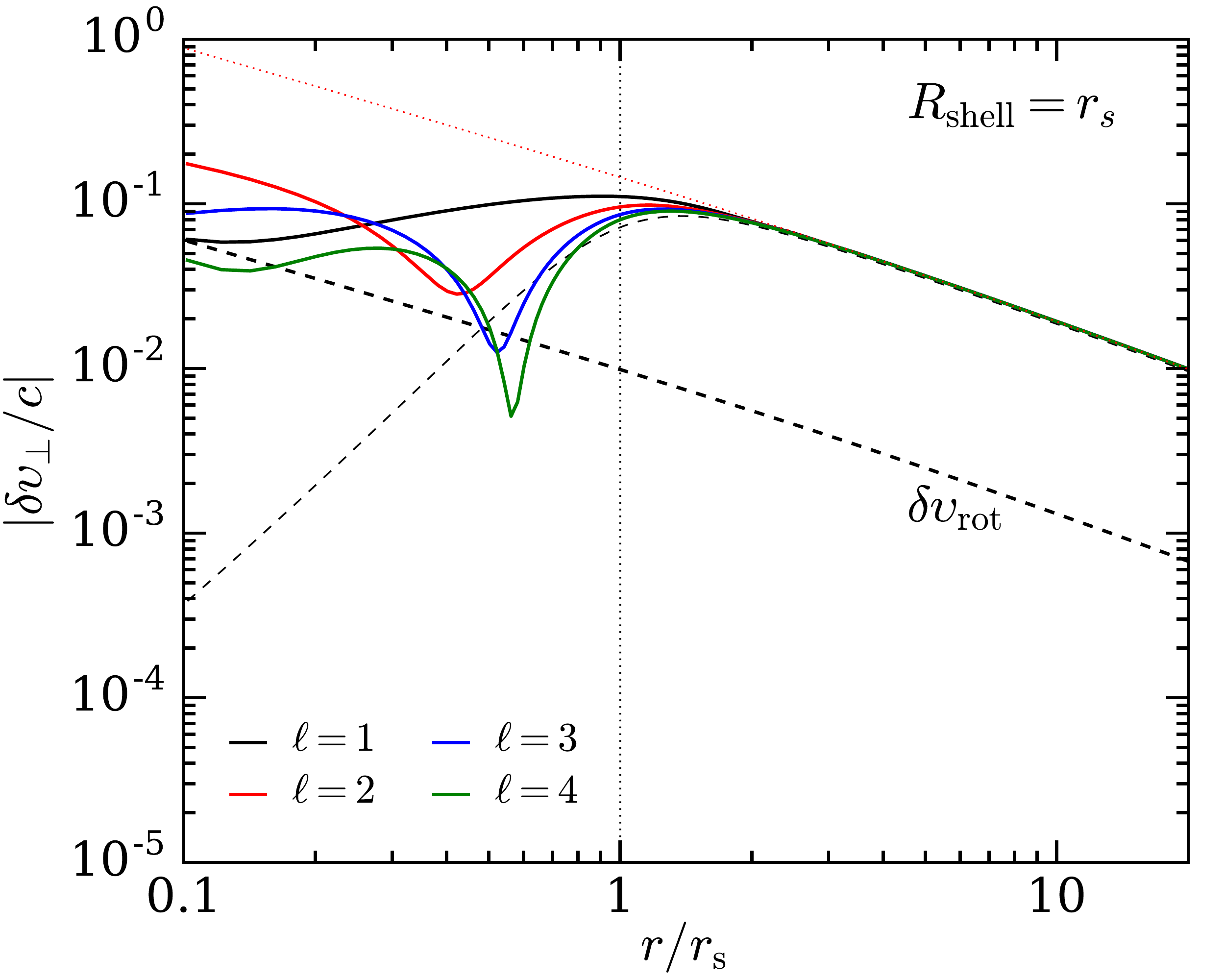}
\includegraphics[angle=0,width=0.65\columnwidth, clip=false]{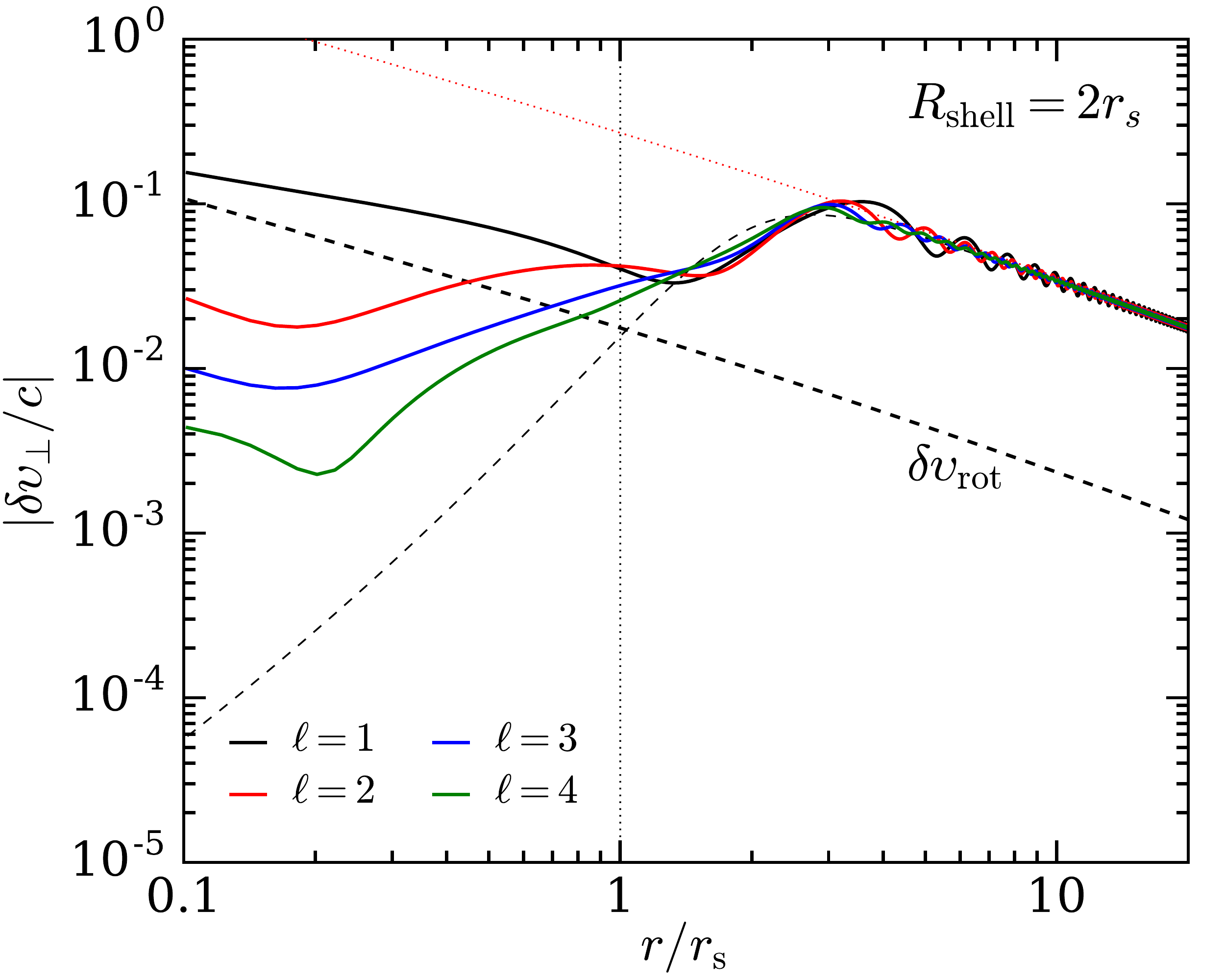}
\includegraphics[angle=0,width=0.65\columnwidth, clip=false]{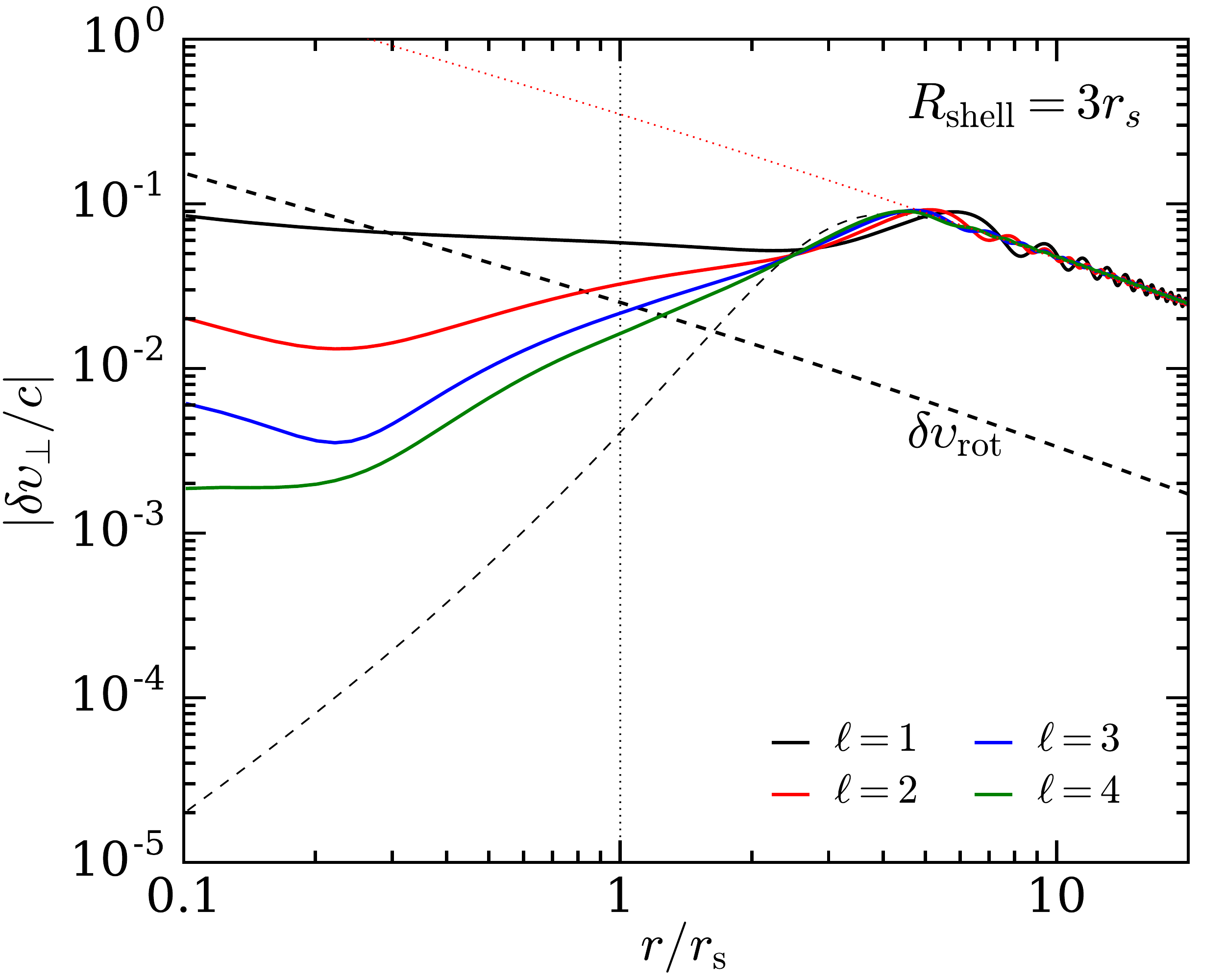}
  \caption{The top panels show the radial profile of the amplitude of $\delta p / \gamma p$ for acoustic waves generated by advected perturbations for different values of angular wavenumber $\ell$ and the initial radius of convective vortices $R_\mathrm{shell}$, deduced from Eq.~(\ref{defgp}). The amplitude $\delta p / \gamma p$ differs drastically at large radii for large $R_\mathrm{shell}$, which is caused by the refraction of acoustic waves. {Middle panels:} Radial profile of $\delta \upsilon_r/c$ generated by advected horizontal vorticity waves for different values of $\ell$ and $R_\mathrm{shell}$, deduced from Eq.~(\ref{deffvr}). {Bottom panels:} Radial profile of $\delta \upsilon_\bot/c$ generated by advected horizontal vorticity waves for different values of $\ell$ and $R_\mathrm{shell}$, deduced from Eq.~(\ref{deffvt}). The dashed lines in the bottom six panel show the contribution of the advected horizontal vorticity waves only (i.e., the contribution of the acoustic waves are excluded), while the solid lines contain the contribution of both acoustic waves and advected horizontal vorticity waves. The straight red dotted lines in all panels show the $\propto r^{-1}$ slope for reference, while the dotted vertical line shows the location of the sonic point. 
   \label{fig:p_vs_r_chi_l}}
\end{center}
\end{figure*}

The radial profiles of the pressure and velocity perturbations generated by the advected entropy and horizontal vorticity perturbations are analyzed for different values of $\ell$ and $R_\mathrm{shell}$ in Fig.~\ref{fig:p_vs_r_chi_l}. The top three panels show the radial profile of $|\delta p/\gamma p|$ for $R_\mathrm{shell}=\rso$, $R_\mathrm{shell}=2\rso$, and $R_\mathrm{shell}=3\rso$. The solid lines with different colors correspond to different values of $\ell$ ranging from 1 till 4. Similarly to the perturbations with $\ell=2$ discussed above (cf. Fig~\ref{fig:p_fudicial}), inside the sonic radius, $|\delta p/\gamma p|$ does not grow much with decreasing $r$. Instead, it exhibits a wave-like pattern with pressure perturbations varying by factor of $\sim 2$. In the supersonic region, the overall value of $|\delta p/\gamma p|$ is consistent with the behavior at radius $0.1 \rso$ that we saw in Fig~\ref{fig:p_all}. At large radii $r>\rso$, $|\delta p/\gamma p|$ is stronger for perturbations with large $R_\mathrm{shell}$ (e.g., $R_\mathrm{shell}\gtrsim 2\rso$). This is caused by the fact that, at large $R_\mathrm{shell}$, perturbations have a large radial size, for which a significant fraction of incoming acoustic waves gets refracted back \citep{foglizzo:01}. These outgoing waves are identified owing to their $\propto r^{-1}$ scaling, which is a consequence of the conservation of energy. Since the amount of refraction decreases with the radial size, relatively little acoustic waves are present for, e.g., $R_\mathrm{shell} = \rso$ at $r\gtrsim \rso$. 

The radial profiles of $\delta \upsilon_r/c$ and $\delta \upsilon_\bot/c$ are shown in the bottom six panels of Fig.~\ref{fig:p_vs_r_chi_l} for the same three values of $R_\mathrm{shell}$. At large radii outside of the sonic point, both components of the velocity increase gradually with decreasing radius. However, this increase saturates around the sonic point and almost no growth takes place in the supersonic region. This is again caused by the stretching of the vortex sheets due to the acceleration of the flow in the radial direction. We again see significantly more refracted acoustic waves at large radii for perturbations with larger $R_\mathrm{shell}$. For outgoing waves, we observe the $ \propto r^{-1}$ scaling which is again a consequence of the conservation of energy. 

It is interesting to compare the total velocity perturbations (i.e., including the contributions of both acoustic and horizontal vorticity perturbations) to the velocity field of only horizontal vorticity waves (i.e., without including the contribution of acoustic waves) in the inner regions of the flow ($r\lesssim R_\mathrm{shell}$). The former is shown with solid lines while the latter is shown with thin dashed lines in the six bottom panels of Fig.~\ref{fig:p_vs_r_chi_l}. The radial velocity acoustic waves is larger than that of horizontal vorticity waves by a factor of a few. However, for $\delta \upsilon_\bot/c$, the contribution of acoustic waves exceed that of horizontal vorticity waves by a factor of $\sim 10^2$ for most perturbation parameters. Thus, the non-radial velocity perturbations ahead of the supernova shock is expected to be dominated by the contribution of acoustic waves. On the other hand, as discussed above, the velocity of the vertical vorticity $\delta \upsilon_\mathrm{rot}$, shown with thick dashed lines in the bottom panels, grows as $\propto 1/r$. If it has an amplitude of $10^{-2}$ at the initial radius $R_\mathrm{shell}$, which is $10$ times smaller than that of the horizontal component, the  contribution of the latter becomes significant. For $R_\mathrm{shell} \ge 3\rso$, $\delta \upsilon_\mathrm{rot}$ becomes the dominant component of the velocity perturbations ahead of the CCSN shock.

 \begin{table}
 \caption{The amplitudes of pressure and velocity perturbations generated by advected entropy and horizontal vorticity waves at radius $0.1\rso$, a radius where CCSN shock is expected to encounter the pre-collapse perturbations. The horizontal vorticity are normalized to yield a Mach number of $0.1$ at the initial radius $R_\mathrm{shell}$, while entropy fluctuations are assumed to have amplitude of $0.05\,k_\mathrm{B}/\mathrm{baryon}$. The vertical component of vorticity scale as $\delta \upsilon_\mathrm{rot}/c \sim 0.06 (R_\mathrm{shell}/\rso)^{0.9}$.}
 \label{tab:models}
 \begin{tabular}{lccccccc}
 \hline
  $\ell$ & $R_\mathrm{shell}$ & $\frac{\delta p^{\mathrm{(v)}}}{\gamma p}$ & $\frac{\delta \upsilon_r^{\mathrm{(v)}}}{c}$ & $\frac{\delta \upsilon_\bot^{\mathrm{(v)}}}{c}$ & $\frac{\delta p^{\mathrm{(e)}}}{\gamma p}$ & $\frac{\delta \upsilon_r^{\mathrm{(e)}}}{c}$ & $\frac{\delta \upsilon_\bot^{\mathrm{(e)}}}{c}$ \\
    & $[\rso]$ & $[10^{-2}]$ & $[10^{-2}]$ & $[10^{-2}]$ & $[10^{-2}]$ & $[10^{-2}]$ & $[10^{-2}]$ \\
\hline
1  &  1  &  17.3  &  4.54  &  5.74  &  0.92  &  0.85  &  2.01 \\ 
1  &  2  &  10.8  &  4.17  &  15.4  &  0.49  &  0.68  &  0.84 \\ 
1  &  3  &  9.64  &  3.45  &  8.22  &  0.59  &  1.06  &  1.89 \\ 
1  &  4  &  8.34  &  2.91  &  5.48  &  1.20  &  1.28  &  2.33 \\ 
2  &  1  &  7.19  &  2.47  &  17.5  &  1.32  &  0.62  &  0.70 \\ 
2  &  2  &  3.39  &  1.86  &  2.53  &  0.45  &  0.97  &  0.77 \\ 
2  &  3  &  1.87  &  1.01  &  1.90  &  0.53  &  1.04  &  0.64 \\ 
2  &  4  &  1.08  &  0.59  &  1.51  &  0.57  &  1.08  &  0.53 \\ 
3  &  1  &  10.9  &  4.83  &  8.73  &  0.67  &  0.54  &  0.76 \\ 
3  &  2  &  1.12  &  1.01  &  0.94  &  0.21  &  0.90  &  0.33 \\ 
3  &  3  &  0.34  &  0.45  &  0.56  &  0.23  &  0.95  &  0.25 \\ 
3  &  4  &  0.16  &  0.27  &  0.31  &  0.21  &  0.96  &  0.22 \\ 
4  &  1  &  6.41  &  3.67  &  4.48  &  0.44  &  0.77  &  0.78 \\ 
4  &  2  &  0.16  &  0.74  &  0.39  &  0.15  &  0.88  &  0.20 \\ 
4  &  3  &  0.02  &  0.44  &  0.08  &  0.12  &  0.91  &  0.17 \\ 
4  &  4  &  0.02  &  0.31  &  0.05  &  0.11  &  0.93  &  0.13 \\ 
5  &  1  &  2.80  &  1.90  &  3.89  &  0.43  &  0.87  &  0.48 \\ 
5  &  2  &  0.06  &  0.77  &  0.04  &  0.08  &  0.86  &  0.13 \\ 
5  &  3  &  0.10  &  0.48  &  0.03  &  0.04  &  0.89  &  0.14 \\ 
5  &  4  &  0.19  &  0.36  &  0.08  &  0.08  &  0.88  &  0.18 \\ 
6  &  1  &  2.01  &  1.12  &  2.08  &  0.31  &  0.81  &  0.19 \\ 
6  &  2  &  0.21  &  0.84  &  0.11  &  0.03  &  0.83  &  0.14 \\ 
6  &  3  &  0.11  &  0.48  &  0.09  &  0.03  &  0.88  &  0.19 \\ 
6  &  4  &  0.05  &  0.31  &  0.08  &  0.06  &  0.91  &  0.20 \\ 
7  &  1  &  1.33  &  1.68  &  0.97  &  0.24  &  0.70  &  0.12 \\ 
7  &  2  &  0.19  &  0.81  &  0.24  &  0.06  &  0.84  &  0.23 \\ 
7  &  3  &  0.04  &  0.45  &  0.12  &  0.06  &  0.89  &  0.19 \\ 
7  &  4  &  0.09  &  0.31  &  0.15  &  0.10  &  0.91  &  0.26 \\ 
8  &  1  &  1.00  &  2.01  &  0.82  &  0.22  &  0.68  &  0.33 \\ 
8  &  2  &  0.08  &  0.75  &  0.25  &  0.09  &  0.86  &  0.22 \\ 
8  &  3  &  0.04  &  0.45  &  0.07  &  0.05  &  0.89  &  0.15 \\ 
8  &  4  &  0.10  &  0.33  &  0.12  &  0.07  &  0.89  &  0.21 \\ 
 	\hline
 \end{tabular}
   \begin{tablenotes}
    \item ${}^\mathrm{(v)}$ Generated by advected vorticity waves.
    \item ${}^\mathrm{(e)}$ Generated by advected entropy waves.
  \end{tablenotes}
\end{table}

\section{Conclusion}
\label{sec:conclusion}
In this work, we have studied the hydrodynamic evolution of convective perturbations in the nuclear-burning shells of massive stars during stellar collapse. The main aim was to investigate the physical properties of the perturbations when they reach the radius of $\sim 150\,\mathrm{km}$, where they are expected to encounter the supernova shock launched at core bounce. The properties of these perturbations affects the way they interact with the shock and thus influence the explosion dynamics. We modeled convection as a combination of vorticity and entropy waves and studied their evolution using linear hydrodynamics equations. Using the transonic Bondi solution to model the collapsing star, we followed the evolution of the hydrodynamic perturbations from large radii at a few $\sim \! 10^3$ km where they originate, down to small radii of $\sim 150\,\mathrm{km}$, where the flow is supersonic. 

As the star collapses, vorticity and entropy perturbations move towards the center together with the stellar matter. Due to the converging geometry of the flow, the horizontal motions contract in the lateral direction, resulting in $\propto r^{-1}$ scaling of the velocity perturbations associated to the vertical vorticity. On the other hand, the velocities associated to the horizontal vorticity decrease with radius. This is caused by the acceleration of the collapse, which stretches the vortex sheets in the radial direction. In order to conserve the circulation, the velocity of vortex sheets has to decrease (cf. Section~\ref{sec:vorticity}). The perturbations with large size are more prone to the radial stretching, leading to smaller velocity perturbations. As a result, ahead of the shock, the Mach number of vorticity waves do not exceed $\sim 0.1$ for most of the perturbation parameters. 

Both entropy and horizontal vorticity perturbations, when advected with the flow, generate acoustic waves (cf. Section~\ref{sec:acoustic_waves}). This happens because, in converging flows, the advected perturbations do not remain in pressure equilibrium. The resulting pressure perturbations propagate as acoustic waves. We find that for models with $\ell=1$ ad $\ell=2$, the pressure perturbations reach the relative amplitude of $\delta p /\gamma p \sim 0.1$ before encountering the supernova shock. This is in agreement with the results of 3D numerical simulations \citep{mueller:17}. The pressure perturbations are smaller for modes with larger $\ell$. Due to the larger radial stretching by accelerated advection, convective vortices with large initial radii generate weak perturbations with a relative amplitude of $\sim 10^{-2}$ or smaller. We find that most of the radial velocity perturbations ahead of the CCSN shock consists of contributions from acoustic and, to a lesser extent, horizontal vorticity perturbations. The non-radial motion is dominated by the contributions from the acoustic waves as well as vertical vorticity perturbations.

Our present work sheds light on the physical properties of the perturbations ahead of the supernova shock. The interaction of 
vorticity, entropy, and acoustic waves with the shock can now be studied in more detail using, e.g., linear theory similar to that of \citet{abdikamalov18} with parameters appropriate for core-collapse supernovae. This will allow us to assess the relative importance of pre-shock acoustic, entropy, and vorticity perturbations on the post-shock dynamics. This will be the subject of a future work. 

Finally, we point out that our work suffers from a number of limitations due to the simplifying assumptions that we employ in our model. In particular, we use the stationary transonic Bondi solution to model the collapsing star. While this solution nicely captures the presence of subsonic and supersonic regions of accretion, in realistic stars, the dynamical collapse proceeds in a non-stationary fashion. Moreover, the real stars contain regions with different compositions and entropies. While these effects are incorporated in recent 3D simulations \cite[e.g.][]{mueller:17}, the Bondi solution does not include them. In addition, the nuclear burning and neutrino cooling may still be taking place during the collapse phase. The adiabatic approximation that we use does not capture these effects. While these simplifications allowed us to obtain a unique insight into the evolution of the perturbations during collapse, the impact of these approximations needs to be carefully evaluated in future works.

\section*{Acknowledgements}

The authors acknowledge fruitful discussions with C\'esar Huete, Ilya Kovalenko, Bernhard M\"uller, Jim Fuller, and Olzhas Mukazhanov. We also thank the anonymous referee for many constructive suggestions. The work was supported by Nazarbayev University Faculty Development Competitive Research Grant No. 090118FD5348, by the Ministry of Education of Kazakhstan's target program IRN: BR05236454 and grant AP05135753. TF benefited from the KITP program on the "Mysteries and inner working of massive stars" supported by the National Science Foundation under Grant No. NSF PHY17-48958.

\bibliographystyle{mnras.bst}

\appendix

\section{Linearized equations for perturbations}
\label{sec:hydro_lin}

We start with the Euler equation,
\begin{equation}
{\p \boldsymbol{\upsilon} \over\p t}+\boldsymbol{\omega}\times \boldsymbol{\upsilon} + \boldsymbol{\nabla} \left({\upsilon^2\over2}+{c^2\over\gamma-1}
-{GM\over r}\right)=c^2 \boldsymbol{\nabla} {S\over\gamma},\label{Euler}
\end{equation}
where $\boldsymbol{\omega}\equiv\nabla \times \boldsymbol{\upsilon}$ is the vorticity vector. The dimensionless entropy $S$ is related to entropy per nucleon via equation $dS = ds_\mathrm{b} / k_\mathrm{b}$, where $k_\mathrm{b}$ is the Boltzmann constant (see Appendix \ref{sec:entropy} for the derivation).
The equation for vorticity $\boldsymbol{\omega}$ can be obtained by combining the curl of Eq.~(\ref{Euler}) with the continuity equation:
\begin{equation}
\frac{\partial }{\partial t} \frac{\boldsymbol{\omega}}{\rho} + 
\left(\boldsymbol{\upsilon} \cdot \boldsymbol{\nabla} \right) \frac{\boldsymbol{\omega} }{\rho} =
\left(\frac{\boldsymbol{\omega}}{\rho} \cdot \boldsymbol{\nabla} \right) \boldsymbol{\upsilon} + \frac{1}{\rho} \boldsymbol{\nabla} c^2 \times 
\boldsymbol{\nabla} \frac{S}{\gamma} \label{eq:vort}
\end{equation}
The projection of the Euler equation along the direction of the flow yields an equation for the Bernoulli constant:
\begin{equation}
\left( \frac{\partial }{\partial t} + \boldsymbol{\upsilon} \cdot \boldsymbol{\nabla} \right) 
\left( \frac{\upsilon^2}{2} + \frac{c^2}{\gamma-1}-\frac{GM}{r} \right) =
\frac{1}{\rho} \frac{\partial p}{\partial t}. \label{dber}
\end{equation}
In the following, we separate the time dependence using the Fourier transform in time. We use the spherical coordinates $(r,\theta,\phi)$ to describe the spatial dependence.
The conservation of entropy during advection implies that
\begin{equation}
\delta S = \delta S_R\eiwv,
\end{equation}
while the conservation of $\delta K$ yields
\begin{equation}
\delta K = \delta K_R\eiwv,
\end{equation}
where $R$ is a coordinate where perturbations have zero phase and $\omega$ is the angular frequency. For clarity, we shall use a prime to distinguish the reference radius $R'$ of the phase of advected perturbation in the supersonic region: $R>\rso$ and $R'<\rso$.
The conservation laws of $\delta K$ and $\delta S$ across the sonic radius relate the solution defined for $R>\rso$ and the solution defined for $R'<\rso$:
\begin{eqnarray}
\label{eq:krprint}
\delta K_{R'}&=&
\delta K_{R}\e^{i\omega\int_{R}^{R'}{\d r\over \upsilon}},\\
\label{eq:srprint}
\delta S_{R'}&=&
\delta S_{R}\e^{i\omega\int_{R}^{R'}{\d r\over \upsilon}}.
\end{eqnarray}
Following \citet{foglizzo:01}, we reformulate the linearized Euler equation using functions $\delta f$ and $\delta g$:
\begin{eqnarray}
\delta f&\equiv& \upsilon\;\delta \upsilon_r + {2\over\gamma-1} c\;\delta c,\label{deff}\\
\delta g&\equiv& {\delta \upsilon_r\over \upsilon} + {2\over\gamma-1}{\delta c\over 
c}.\label{defg}
\end{eqnarray}
The perturbations of the hydrodynamics quantities such as $\delta \upsilon_r$, $\delta c$, $\delta \rho$ and $\delta p$ corresponding to $\delta f$ and $\delta g$ can be obtained by simply inverting relations (\ref{deff})-(\ref{defg}) \citep{foglizzo:07}:
\begin{eqnarray}
\frac{\delta \upsilon_r}{\upsilon} &=& \frac{1}{1-{\cal M}^2} \left(\delta g-\frac{\delta f}{c^2}\right),\qquad\label{deffvr}\\
\frac{\delta c^2}{c^2} &=& \frac{\gamma-1}{1-{\cal M}^2} \left(\frac{\delta f}{c^2} - {\cal M}^2 \delta g\right),\qquad\label{defgc}\\
\frac{\delta \rho}{\rho} &=& \frac{1}{1-{\cal M}^2} \left( -{\cal M}^2 \delta g - (1-{\cal M}^2) \delta S + \frac{\delta f}{c^2} \right),\qquad\label{defgrho}\\
\frac{\delta p}{\gamma p} &=& \frac{1}{1-{\cal M}^2} \left( -{\cal M}^2 \delta g - (1-{\cal M}^2) \frac{\delta S}{\gamma} + \frac{\delta f}{c^2} \right).\qquad \label{defgp}
\end{eqnarray}
The transverse velocity component can be expressed in terms of $\delta f$ and $\delta K$ (cf. Appendix~\ref{sec:deltaK}): 
\begin{equation}
\delta \upsilon_\bot = \frac{1}{i \omega r} \left( \delta f - \frac{\delta K}{L^2} \right). \label{deffvt}
\end{equation}
We can obtain a system of differential equations for $\delta f$ and $\delta g$ by combining the continuity equation with the radial projection of the Euler equation:
\begin{eqnarray}
\upsilon{\p \delta f\over\p r}+{i\omega \M^2\delta f\over 1-\M^2}
&= &
{i\omega \upsilon^2 \delta g\over 1-\M^2} + i\omega c^2 {\delta 
S_R\over\gamma}\eiwv,\qquad \\
\upsilon{\p \delta g\over\p r}+{i\omega \M^2\delta g\over 1-\M^2}
&=& {i\omega \delta f\over c^2(1-\M^2)} \nonumber \\ 
&+& 
{i\over \omega}\Delta_{\theta,\varphi}\delta f+{i\delta K_R\over r^2\omega 
}\eiwv,
\end{eqnarray}
where $\Delta_{\theta,\varphi}$ is the angular part of the Laplacian. The homogeneous system associated with this system describes propagation of free acoustic waves. In the presence of inhomogeneous terms $ \delta K$ and $\delta S$, which model advected vorticity and entropy perturbations, the solution of this system has multiple components: the vorticity and entropy perturbations themselves as well as the acoustic waves that these two perturbations generate. The contribution of acoustic waves as well as vorticity and entropy waves to the values of $\delta f$ and $\delta g$ can be separated using the decomposition of \citet{foglizzo:07}, as described in Appendix~\ref{sec:decomposition}.
Using the spherical harmonics $Y_l^m(\theta,\varphi)$ decomposition, we obtain:
\begin{eqnarray}
\label{dfdr}
\upsilon{\p \delta f\over\p r}+{i\omega \M^2\delta f\over 1-\M^2}
&=& {i\omega \upsilon^2 \delta g\over 1-\M^2} + i\omega c^2 {\delta 
S_R\over\gamma}\eiwv,\qquad
\\
\label{dgdr}
\upsilon{\p \delta g\over\p r}+{i\omega \M^2\delta g\over 1-\M^2}
&=& 
{i\omega \delta f\over c^2(1-\M^2)} - {iL^{2}\over \omega r^2} \delta f \nonumber \\ 
&+& 
{i\delta K_R\over r^2\omega }\eiwv.
\end{eqnarray}
In either region $r>\rso$ or $r<\rso$, we define quantities $\delta \tilde f$ and $\delta \tilde g$ as:
\begin{eqnarray}
{\delta \tilde f}&\equiv&\e^{i\omega\int_R^r{\M^2\over1-\M^2}{\d r\over \upsilon}}\delta f,\label{ftilde}\\
{\delta \tilde g}&\equiv&\e^{i\omega\int_R^r{\M^2\over1-\M^2}{\d r\over \upsilon}}\delta g.
\end{eqnarray}
where the lower bound $R$ of the integral is chosen in the same region. Despite the mathematical singularity at $r=\rso$, the differential system deduced from equations~(\ref{dfdr})-(\ref{dgdr}) in each half domain $r>\rso$ or $r<\rso$ is formally simpler:
\begin{eqnarray}
{\p \delta \tilde f\over\p r}&=& {i\omega \upsilon {\delta \tilde g}\over1-\M^2} 
+ i\omega{c^2\over \upsilon}
{\delta S_R\over\gamma}\e^{i\omega\int_R^r {\d r\over \upsilon(1-\M^2)}},
\label{dtfdr}\\
{\p \delta \tilde g\over\p r}&=& {i{\delta \tilde f}\over \omega \upsilon}
\left\lbrack{\omega^2\over c^2(1-\M^2)} - {L^{2}\over 
r^2}\right\rbrack
+{i\delta K_R\over r^2\omega \upsilon}
\e^{i\omega\int_R^r {\d r\over \upsilon(1-\M^2)}}.\hspace{0.7cm}
\label{dtgdr}
\end{eqnarray}
Using the new variable $X$, which is related to $r$ via equation
\begin{equation}
\label{eq:x}
{\d X\over\d r} \equiv {\upsilon\over 1-\M^2},
\end{equation}
system (\ref{dtfdr})-(\ref{dtgdr}) can be combined into a more compact form:
\begin{eqnarray}
\label{canonic}
{\p^2 \delta \tilde f\over\p X^2}+W \tilde f 
&=&
-{1-\M^2\over \upsilon}\e^{i\omega\int_R^r{\d X\over \upsilon^2}}
\nonumber \\ 
&\times &
\left\lbrace
{\omega\over \M^2}{\delta S_R\over \gamma}\left( {\omega\over \upsilon}
+i{\p\log\M^2\over\p r}\right)+{\delta K_R\over \upsilon r^2}
\right\rbrace, \nonumber \\
\end{eqnarray}
where 
\begin{equation}
W \equiv {1\over \upsilon^2c^{2}}(\omega^2-\omega_{l}^{2}). \label{defW}
\end{equation}
and
\begin{equation}
\omega_{l}^2\equiv l(l+1){c^2-\upsilon^2\over r^2}.\label{omegal}
\end{equation}

\section{Approximate solutions of the homogeneous equation}
\label{sec:homogeneous_solution}

\subsection{WKB approximation at large radii\label{Awkb}}

\begin{figure}
\begin{center}
\includegraphics[angle=0,width=\columnwidth,clip=false]{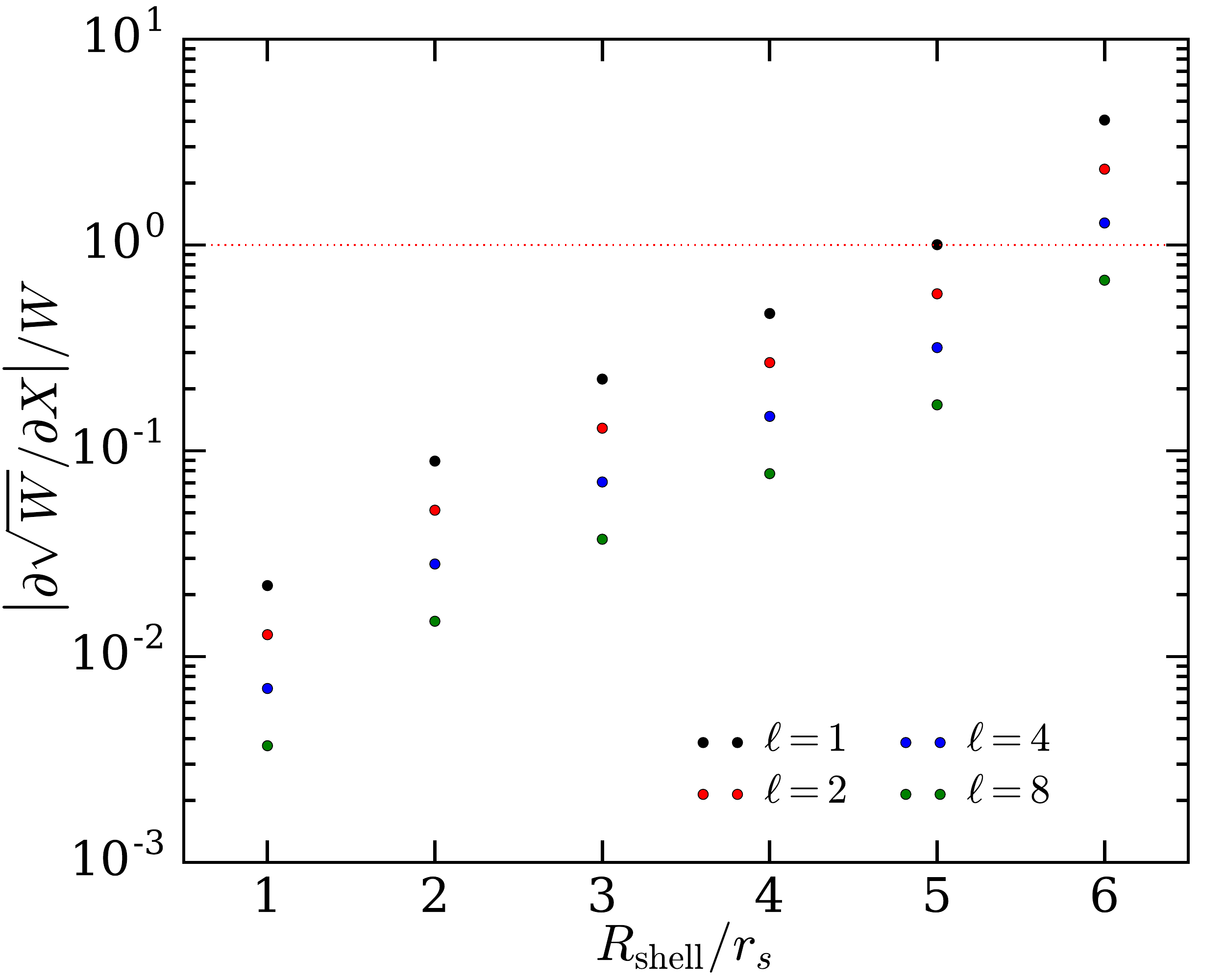}
  \caption{Ratio $|\p \sqrt{W} / \p X | / W$, which measures the degree of the validity of the WKB approximation, as a function of initial radius of the vortices for different values of $\ell$ at radius $r=40\rso$. 
   \label{fig:wkb}}
\end{center}
\end{figure}

The general solution of the homogeneous equation, 
\begin{equation}
\label{canonic2}
{\p^2 \delta  \tilde f\over\p X^2}+W \delta \tilde f = 0,
\end{equation}
associated with equation~(\ref{canonic}) is a linear combination of outgoing ($\delta f^-$) and ingoing ($\delta f^+$) acoustic waves. At the outer boundary, we obtain these two using the WKB approximation \citep{foglizzo:01}:
\begin{eqnarray}
{\delta \tilde f^\pm}\sim A_\pm{\omega^{1\over2}\over W^{1\over4}}
\e^{i\omega\int_R^\infty{\M^2\over1-\M^2}{\d r\over \upsilon}}
\exp \left(\pm i\int^r {\upsilon W^{1\over2}\over 1-\M^2}\d r\right),
\label{defsonadiab}\\
{\delta f^\pm}\sim A_\pm{\omega^{1\over2}\over W^{1\over4}}
\exp \left(i\omega\int_r^\infty{\M^2\over1-\M^2}{\d r\over \upsilon}\pm i\int^r {\upsilon W^{1\over2}\over 1-\M^2}\d r\right),
\end{eqnarray}
where $A_\pm$ is a complex amplitude such that $|A_-|=|A_+|$ is homogeneous to a velocity. The WKB approximation is satisfied at large radii from the center or for high-frequency perturbations. These two conditions are consistent with the requirement that 
\begin{eqnarray}
\left| {\p \sqrt{W} \over\p X} \right| \ll W,\label{wkbhigh}
\end{eqnarray}
Figure~\ref{fig:wkb} shows the ratio $|\p \sqrt{W} / \p X | / W$, which measures the degree of the validity of the WKB approximation, as a function of initial radius of the vortices $R_\mathrm{shell}$ for different values of $\ell$ at the outer boundary of our computational domain. The latter is chosen to be at $r=40\rso$ for our setup. As we can see the ratio is below $1$ for $R_\mathrm{shell} < 5 \rso$. For this reason, we consider initial radii ranging from $\rso$ to $4\rso$, where WKB is expected to yield accurate result.

The Wronskien ${\cal W}$ of ${\delta \tilde f^+}$ and ${\delta \tilde f^-}$ (or the pair of solutions ${\delta \tilde f_0}$ and ${\delta \tilde f^-}$), on either side of the sonic point, is:
\begin{eqnarray}
{\cal W}&\equiv&{\tilde f^+}{\p {\tilde f^-}\over\p r}-
{\tilde f^-}{\p{\tilde f^+}\over\p r}
= -{2i\omega \upsilon\over1-\M^2}A_R,\label{normwronskien}\\
A_R&\equiv&A_+A_-
\e^{2i\omega\int_R^\infty{\M^2\over1-\M^2}{\d r\over \upsilon}}
.
\end{eqnarray}
The Wronskien of $(\delta f_0,\delta f^-)$ or $(\delta f^+,\delta f^-)$ is independent of the boundary $R$:
\begin{eqnarray}
{\delta f_0}{\p {\delta f^-}\over\p r}-
{\delta f^-}{\p{\delta f_0}\over\p r}
&=&-{2i\omega \upsilon\over1-\M^2}A_+A_-\e^{-2i\omega\int_\infty^r{\M^2\over1-\M^2}{\d r\over \upsilon}}. \nonumber\\
\end{eqnarray}
We note that $\delta \tilde f_0$ is singular at the sonic point. On either side of the sonic radius, 
\begin{eqnarray}
{\delta \tilde f^-}{\delta \tilde g_0}-{\delta \tilde f_0}{\delta \tilde g^-}&=&2
A_R
,\label{normwronskienfg}\\
{\delta f^-}{\delta g_0}-{\delta f_0}{\delta g^-}&=&2
A_+A_-
\e^{-2i\omega\int_\infty^r{\M^2\over1-\M^2}{\d r\over 
\upsilon}}.
\end{eqnarray}

\subsection{Approximation in the supersonic region}

At high Mach number the velocity approaches free fall and the sound speed is deduced from mass conservation of the isentropic gas:
\begin{eqnarray}
\upsilon&\propto& r^{-{1\over 2}},\\
c&\propto&\left({1\over \upsilon r^2}\right)^{\gamma-1\over 2}\sim r^{-{3\over 4}(\gamma-1)},\\
\M&\propto&r^{{3\over 4}(\gamma-1)-{1\over 2}}
\end{eqnarray}
The phase relation between $\delta f$ and $\delta \tilde f$ is thus a converging function when $r\to 0$.
According to the differential system (\ref{canonic}),
\begin{eqnarray}
{\p^2 \delta f\over \p r^2}\propto {\delta f\over r^2c^2}\propto \delta f r^{-{3\over 2}}
\end{eqnarray}
It implies that the homogeneous solution $\delta f$ is bounded when $r\to 0$.
\begin{eqnarray}
\delta f\propto \e^{r^{{1\over 2}}}
\end{eqnarray}

\section{Solutions with entropy and vorticity perturbations}
\label{sec:acoustic}

\subsection{Solution for vorticity perturbations}
\label{sec:solution_vorticity}

The solution of equation (\ref{canonic}) for the case with $\delta K \ne 0$ and $\delta S = 0$ 
can be obtained using the method of variation of parameters. The two free parameters of the 
method are fixed by (1) imposing the regularity at $r=\rso$ and (2) assuming that no sound waves come 
from infinity, which leads to the solution \citep{foglizzo:01} 
\begin{eqnarray}
\delta f(r>\rso)&=&-{i\delta K_R\over2\omega A_R} 
\nonumber \\ 
&\times& \bigg\{ \delta f^-
\int_{\rso}^r
\e^{i\omega\int_R^r{1+\M^2\over 1-\M^2}{\d r\over \upsilon}}
{\delta f_0\over r^2\upsilon}
\d r 
\nonumber \\ 
&-&\delta f_0
\int_{\infty}^r
\e^{i\omega\int_R^r{1+\M^2\over 1-\M^2}{\d r\over \upsilon}}
{\delta f^-\over r^2\upsilon}
\d r
\bigg\},\label{outsidem}
\end{eqnarray}
where $R>\rso$, $\delta f_0$ is the regular homogeneous solution. $\delta f^-$ corresponds to outgoing acoustic waves when $r\gg\rso$, normalized according to Eq.~(\ref{defsonadiab}). The function $\delta f^-$ is singular at the sonic radius.
The Wronskien associated to the pair ($\delta f_0,\delta f^-)$ satisfies Eq.~(\ref{normwronskien}).\\
As in \citet{foglizzo:01}, an integration by part is used to accelerate the convergence as $r^{-5}$ at infinity
\begin{eqnarray}
&&{\delta f}(r>\rso)={\delta K_R\over2\omega^2A_R} \times \nonumber \\ &&
\bigg\{
{\delta f^-}
\int_{r_{\rm s}}^{r}
\e^{i\omega\int_{R}^r {1+\M^2\over1-\M^2}{\d r\over \upsilon}}
\left\lbrack
{\p\over\p r}\left({1-\M^{2}\over r^{2}}\right)\delta f_0
+{i\omega \upsilon\over r^{2}}\delta g_0
\right\rbrack\d r \nonumber\\
&&-{\delta f_0}
\int_{\infty}^{r}
\e^{i\omega\int_{R}^r {1+\M^2\over1-\M^2}{\d r\over \upsilon}}
\left\lbrack
{\p\over\p r}\left({1-\M^{2}\over r^{2}}\right)\delta f^-
+{i\omega \upsilon\over r^{2}}g^-
\right\rbrack\d r
\bigg\},\nonumber \\ \label{outsidemnum}
\end{eqnarray}
We use the regular solution $\delta f_0$ and the technique of variation of constants to define a second solution $\delta f_{\rm sup}$ of the homogeneous equation in the supersonic region:
\begin{eqnarray}
\delta f_{\rm sup}(r<\rso) &\equiv&
-2i\omega A_R \delta f_0 \nonumber \\ 
& \times &
\int_{R'}^r\e^{-2i\omega\int_{R'}^r{\M^2\over1-\M^2}{\d r\over \upsilon}}{\upsilon \over \delta f_0^2}{\d r\over 1-\M^2}, 
\nonumber \\ \label{deffR}
\end{eqnarray}
It is singular at the sonic point. The singularity of the integral is isolated using an integration by parts:
\begin{eqnarray}
\delta f_{\rm sup}(r)&=&  A_R\delta f_0 \bigg\{
\left\lbrack\e^{-2i\omega\int_{R'}^r{\M^2\over1-\M^2}{\d r\over \upsilon}}{c^2\over \delta f_0^2}\right\rbrack_{R'}^r 
\nonumber\\ 
&-& \int_{R'}^r\e^{-2i\omega\int_{R'}^r{\M^2\over1-\M^2}{\d r\over \upsilon}}
{\p\over\p r}\left({c^2\over \delta f_0^2}\right)\d r \bigg\} ,
\nonumber\\ &=&  
A_R\delta f_0 \bigg\{ \left\lbrack\e^{-2i\omega\int_{R'}^r{\M^2\over1-\M^2}{\d r\over \upsilon}}{c^2\over \delta f_0^2}\right\rbrack_{R'}^r \nonumber\\ 
&-& \int_{R'}^r\e^{-2i\omega\int_{R'}^r{\M^2\over1-\M^2}{\d r\over \upsilon}} 
\nonumber \\
&\times&
{1\over \delta f_0^3}\left(
\delta f_0{\p c^2\over\p r}
-2c^2{\p \delta f_0\over\p r}
\right)
\d r \bigg\}.
\end{eqnarray}
The singular phase is also calculated using an integration by parts:
\begin{eqnarray}
\int_{R'}^r{1+\M^2\over1-\M^2}{\d r\over \upsilon}&=&\left\lbrack {1+\M^2\over \upsilon}{r-\rso\over1-\M^2}\log |r-\rso|\right\rbrack_{R'}^r
\nonumber\\ 
&-&\int_{R'}^r\log |r-\rso|{\p\over\p r}\left({1+\M^2\over \upsilon}{r-\rso\over1-\M^2}\right){\d r},
\nonumber\\
\end{eqnarray}
or
\begin{eqnarray}
&& \int_{R'}^r{1+\M^2\over1-\M^2}{\d r\over \upsilon}=
\nonumber \\ &&
-\left\lbrack {1+\M^2\over \upsilon}\left({\p\M^2\over\p r}\right)^{-1}\log |1-\M^2|\right\rbrack_{R'}^r 
\nonumber \\ &&
+\int_{R'}^r\log |1-\M^2|{\p\over\p r}\left\lbrack{1+\M^2\over \upsilon}
\left({\p\M^2\over\p r}\right)^{-1}\right\rbrack{\d r}.
\end{eqnarray}
In derivation of the last equations, we have used the radial derivated of the Mach number:
\begin{equation}
{\p\M^2\over\p r}=
2(\gamma-1){\M^2\over r}-{\gamma+1\over1-\M^2}
\left(2-{1\over rc^2}\right){\M^2\over r},
\end{equation}
and
\begin{eqnarray}
{\p^2\M^2\over\p r^2}&=&
-2(\gamma-1){\M^2\over r^2}
+2{\gamma-1\over r}{\p\M^2\over \p r} \nonumber \\
&&-{\gamma+1\over (1-\M^2)^2}
\left({2\over r}-{1\over r^2c^2}\right)
{\p\M^2\over\p r}
\nonumber \\ &&
+{\gamma+1\over r^2}{2\M^2\over1-\M^2}
\left\lbrack1-{1\over c^2}\left({1\over r}+{\p\log c\over\p r}\right)\right\rbrack.
\end{eqnarray}
The definition of the function $\delta g_{\rm sup}$ follows from Eq.~(\ref{dtgdr}):
\begin{eqnarray}
\delta g_{\rm sup}(r<\rso)\equiv  {1\over \delta f_0}\left(\delta g_0\delta f_{\rm sup}-2 A_R\e^{-2i\omega\int_{R'}^r{\M^2\over1-\M^2}{\d r\over \upsilon}}\right).
\end{eqnarray}
The normalization factor $(-2i\omega A_R)$ in Eq.~(\ref{deffR}) has been chosen so that the Wronskien of ($\delta f_0,\delta f_{\rm sup}$) is the same as ($\delta f_0,\delta f^-$) as defined by Eq.~(\ref{normwronskien}).
We define a general solution in the supersonic part of the flow which is regular at the sonic point and matches the subsonic solution given by Eq.~(\ref{outsidem}) at $r=\rso$:
\begin{eqnarray}
{\delta f}(r<\rso)&=&-{i\delta K_{R'}\over2\omega A_R} \times \nonumber \\ &&
\bigg\{
\delta f_{\rm sup}
\int_{\rso}^r{\delta f_0\over r^2\upsilon}
\e^{i\omega\int_{R'}^r{1+\M^2\over 1-\M^2}{\d r\over \upsilon}}\d r \nonumber\\ &&
-\delta f_0
\int_{\rso}^r{\delta f_{\rm sup}\over r^2\upsilon}
\e^{i\omega\int_{R'}^r{1+\M^2\over 1-\M^2}{\d r\over \upsilon}}\d r \nonumber\\ &&
-{\delta f_0}\e^{i\omega\int_{R'}^{R}{\d r\over \upsilon}}
\!\! \int_{\infty}^{\rso} \! {\delta f^-\over r^2\upsilon}
\e^{i\omega\int_{R}^r{1+\M^2\over 1-\M^2}{\d r\over \upsilon}}
\d r
\bigg\}. 
\qquad\,\,\label{pressure_vorticity}
\end{eqnarray}
A faster convergence near the origin is obtained by using an integration by parts:
\begin{eqnarray}
\int_{\rso}^r 
\e^{i\omega\int_{R'}^r{1+\M^2\over1-\M^2}{\d r\over \upsilon}}
{{\delta f_0}\over r^2\upsilon}
\d r \nonumber \\ 
= \int_{\rso}^r
\e^{i\omega\int_{R'}^r{1\over1-\M^2}{\d r\over \upsilon}}
{i\omega \over L^2+{\omega^2r^2\over \upsilon^2-c^2}}{\p {\delta \tilde g_0}\over \p r}
\d r \nonumber \\ 
= \left\lbrack
\e^{i\omega\int_{R'}^r{1\over1-\M^2}{\d r\over \upsilon}}
{i\omega \over L^2+{\omega^2r^2\over \upsilon^2-c^2}}{\delta \tilde g_0}
\right\rbrack_{\rso}^r
\nonumber \\ 
- \int_{\rso}^r {\delta \tilde g_0}
{\p\over\p r}\left(\e^{i\omega\int_{R'}^r{1\over1-\M^2}{\d r\over \upsilon}}
{i\omega \over L^2+{\omega^2r^2\over \upsilon^2-c^2}}\right)
\d r.
\end{eqnarray}
In consequence, each integral is now convergent when $r\to 0$:
\begin{eqnarray}
&&{\delta f_{\rm sup}}
\int_{\rso}^r{{\delta \tilde f_0}\over r^2 \upsilon} 
\e^{i\omega\int_{R'}^r{\d r\over \upsilon(1-\M^2)}}\d r \nonumber \\ &&
-{\delta f_0}\int_{\rso}^r{{\delta \tilde f_{\rm sup}}\over r^2\upsilon}
\e^{i\omega\int_{R'}^r{\d r\over \upsilon(1-\M^2)}}\d r 
=\e^{i\omega\int_{R'}^r{\d r\over \upsilon}}
{2i\omega A_R \over L^2+{\omega^2r^2\over \upsilon^2-c^2}}
\nonumber \\
&&
-{\delta f_{\rm sup}}\int_{\rso}^r {\delta \tilde g_0}
{\p\over\p r}\left(\e^{i\omega\int_{R'}^r{1\over1-\M^2}{\d r\over \upsilon}}
{i\omega \over L^2+{\omega^2r^2\over \upsilon^2-c^2}}\right)
\d r \nonumber \\ &&
+{\delta f_0}\int_{\rso}^r {\delta \tilde g_{\rm sup}}
{\p\over\p r}\left(\e^{i\omega\int_{R'}^r{1\over1-\M^2}{\d r\over \upsilon}}
{i\omega \over L^2+{\omega^2r^2\over \upsilon^2-c^2}}\right)
\d r .
\end{eqnarray}
Note that, in deriving the last equation, we used the relation $\delta \tilde f_\mathrm{sup} \delta \tilde g_0 - \delta \tilde f_0 \delta \tilde g_\mathrm{sup} = 2A_R$. Using equation (\ref{dtfdr}), we can rewrite this relation as 
\begin{eqnarray}
{\delta f_{\rm sup}}
\int_{\rso}^r{{\delta \tilde f_0}\over r^2\upsilon}
\e^{i\omega\int_{R'}^r{\d r\over \upsilon(1-\M^2)}}\d r 
\nonumber \\ 
-{\delta f_0}\int_{\rso}^r{{\delta \tilde f_{\rm sup}}\over r^2\upsilon}
\e^{i\omega\int_{R'}^r{\d r\over \upsilon(1-\M^2)}}\d r 
=\e^{i\omega\int_{R'}^r{\d r\over \upsilon}}
{2i\omega A_R\over L^2+{\omega^2r^2\over \upsilon^2-c^2}}
\nonumber\\
-{\delta f_{\rm sup}}\int_{\rso}^r {1-\M^2\over  \upsilon}{\p{\delta \tilde f_0}\over\p r}
{\p\over\p r}\left({\e^{i\omega\int_{R'}^r{1\over1-\M^2}{\d r\over \upsilon}}
 \over L^2+{\omega^2r^2\over \upsilon^2-c^2}}\right)
\d r \nonumber \\ 
+{\delta f_0}\int_{\rso}^r {1-\M^2\over \upsilon}{\p{\delta \tilde f_{\rm sup}}\over\p r}
{\p\over\p r}\left({\e^{i\omega\int_{R'}^r{1\over1-\M^2}{\d r\over \upsilon}}
\over L^2+{\omega^2r^2\over \upsilon^2-c^2}}\right)
\d r,  \nonumber 
\end{eqnarray}
thus
\begin{eqnarray}
{\delta f_{\rm sup}}
\int_{\rso}^r{{\delta \tilde f_0}\over r^2\upsilon}
\e^{i\omega\int_{R'}^r{\d r\over \upsilon(1-\M^2)}}\d r 
\nonumber \\ 
-{\delta f_0}\int_{\rso}^r{{\delta \tilde f_{\rm sup}}\over r^2\upsilon}
\e^{i\omega\int_{R'}^r{\d r\over \upsilon(1-\M^2)}}\d r
=\e^{i\omega\int_{R'}^r{\d r\over \upsilon}}
{2i\omega A_R\over L^2+{\omega^2r^2\over \upsilon^2-c^2}}
\nonumber\\
+{\delta f_{\rm sup}}\! \int_{\rso}^r \! {\delta \tilde f_0}{\p\over\p r}\left\lbrack{1-\M^2\over  \upsilon}
{\p\over\p r}\left({\e^{i\omega\int_{R'}^r{1\over1-\M^2}{\d r\over \upsilon}}
 \over L^2+{\omega^2r^2\over \upsilon^2-c^2}}\right)\right\rbrack
\d r
\nonumber\\
-{\delta f_0}\! \int_{\rso}^r \! {\delta \tilde f_{\rm sup}}{\p\over\p r}\left\lbrack{1-\M^2\over  \upsilon}
{\p\over\p r}\left({\e^{i\omega\int_{R'}^r{1\over1-\M^2}{\d r\over \upsilon}}
\over L^2+{\omega^2r^2\over \upsilon^2-c^2}}\right)\right\rbrack
\d r.
\end{eqnarray}

\subsection{Acoustic field of entropy perturbations}
\label{sec:solution_entropy}

The general solution for the advected entropy waves can be obtained by linearly superposing the solution for $\delta K = L^2 \delta S/\gamma$, which accounts for the contribution of the vorticity generated by the advected entropy waves, with the solution for $\delta S \ne 0$ and $\delta K=0$. The latter can be written as follows, provided that it is regular at the sonic point and provided that there are no acoustic waves coming from infinity \citet{foglizzo:01},
\begin{eqnarray}
&&\delta f(r>\rso)=-{\delta S_R\over2\gamma A_R} \nonumber \\ && 
\times \bigg\{ \delta f^-
\int_{\rso}^r
{\delta \tilde f_0}{\p\over\p r}
\left( {1-\M^2\over \M^2}\e^{i\omega\int_R^r{\d r\over \upsilon(1-\M^2)}}
\right)\d r \nonumber \\ &&
-\delta f_0
\int_{\infty}^r
{\delta \tilde f^-}
{\p\over\p r}
\left( {1-\M^2\over \M^2}
\e^{i\omega\int_R^r{\d r\over \upsilon(1-\M^2)}}
\right)
\d r
\bigg\}, \qquad
\end{eqnarray}
After an integration by parts the integrated terms cancel out:
\begin{eqnarray}
&&\delta f(r>\rso)={\delta S_R\over2\gamma A_R} \nonumber \\ &&
\times \bigg\{ \delta f^-
\int_{\rso}^r
{\p{\delta \tilde f_0}\over\p r}
\left( {1-\M^2\over \M^2}\e^{i\omega\int_R^r{\d r\over \upsilon(1-\M^2)}}
\right)\d r \nonumber \\ &&
-\delta f_0
\int_{\infty}^r
{\p{\delta \tilde f^-}\over\p r}
\left( {1-\M^2\over \M^2}\e^{i\omega\int_R^r{\d r\over \upsilon(1-\M^2)}}
\right)
\d r
\bigg\}. \qquad 
\end{eqnarray}
After a second integration by parts, the integrals converge at infinity:
\begin{eqnarray}
&& \delta f(r>\rso)=-{\delta S_R\over2\gamma i\omega A_R} \nonumber \\ &&
\times \bigg\{
- {\upsilon(1-\M^2)^2\over \M^2}
\left(\delta f^-{\p{\delta \tilde f_0}\over\p r}-\delta f_0{\p{\delta \tilde f^-}\over\p r}\right)
\e^{i\omega\int_R^r{\d r\over \upsilon(1-\M^2)}}
 \nonumber\\ && 
+\delta f^-
\int_{\rso}^r
{\p\over\p r}\left\lbrack {\upsilon(1-\M^2)^2\over \M^2}{\p{\delta \tilde f_0}\over\p r}\right\rbrack
\e^{i\omega\int_R^r{\d r\over \upsilon(1-\M^2)}}
\d r  \nonumber\\ && 
-
\delta f_0
\int_{\infty}^r
{\p\over\p r}\left\lbrack {\upsilon(1-\M^2)^2\over \M^2}{\p{\delta \tilde f^-}\over\p r}\right\rbrack
\e^{i\omega\int_R^r{\d r\over \upsilon(1-\M^2)}}
\d r
\bigg\}
,\nonumber
\end{eqnarray}
thus
\begin{eqnarray}
&&\delta f(r>\rso)=
{\delta S_R\over\gamma }(c^2-\upsilon^2)
\e^{i\omega\int_R^r{\d r\over \upsilon}} -{\delta S_R\over2\gamma i\omega A_R}
\nonumber \\
&& \times \bigg\{ \delta f^-
\int_{\rso}^r
{\p\over\p r}\left\lbrack {\upsilon(1-\M^2)^2\over \M^2}{\p{\delta \tilde f_0}\over\p r}\right\rbrack
\e^{i\omega\int_R^r{\d r\over \upsilon(1-\M^2)}}
\d r \nonumber\\ &&
-
\delta f_0
\int_{\infty}^r
{\p\over\p r}\left\lbrack {\upsilon(1-\M^2)^2\over \M^2}{\p{\delta \tilde f^-}\over\p r}\right\rbrack
\e^{i\omega\int_R^r{\d r\over \upsilon(1-\M^2)}}
\d r
\bigg\}, \nonumber
\end{eqnarray}
or
\begin{eqnarray}
&&\delta f(r>\rso)=
{\delta S_R\over\gamma }(c^2-\upsilon^2)
\e^{i\omega\int_R^r{\d r\over \upsilon}}
\nonumber \\
&& -{\delta S_R\over2\gamma A_R}\bigg\{ \delta f^-
\int_{\rso}^r
{\p\over\p r}\left\lbrack (c^2-\upsilon^2){\delta \tilde g_0}\right\rbrack
\e^{i\omega\int_R^r{\d r\over \upsilon(1-\M^2)}}
\d r \nonumber\\ &&
-
\delta f_0
\int_{\infty}^r
{\p\over\p r}\left\lbrack (c^2-\upsilon^2){\delta \tilde g^-}\right\rbrack
\e^{i\omega\int_R^r{\d r\over \upsilon(1-\M^2)}}
\d r
\bigg\},
\end{eqnarray}
The functions $A_{k}(r),B_{k}(r)$ are obtained by integrating by parts:
\begin{eqnarray}
&&\delta f(r>\rso)=
{\delta S_R\over\gamma }
D_k\e^{i\omega\int_R^r{\d r\over \upsilon}} 
\nonumber \\ &&
+ {\delta S_R\over2\gamma A_R} \bigg\{ \delta f^-
\int_{\rso}^r
\e^{i\omega\int_R^r{1+\M^2\over 1-\M^2}{\d r\over \upsilon}}
(A_k \delta f_0+B_k \delta g_0)
\d r \nonumber \\ &&
-
\delta f_0
\int_{\infty}^r
\e^{i\omega\int_R^r{1+\M^2\over 1-\M^2}{\d r\over \upsilon}}
(A_k \delta f^-+B_k \delta g^-)
\d r
\bigg\},
\end{eqnarray}
with
\begin{eqnarray}
A_{1} &\equiv& - {i\omega\over \upsilon}
\left(1-{\omega_{l}^2\over\omega^{2}}\right)\propto r^2,\\
B_{1}&\equiv& -{\p\over\p r}(c^{2}-\upsilon^2)\propto r^{-2},\\
D_{1}&\equiv& c^2-\upsilon^2
\end{eqnarray}
After an integration by parts
\begin{eqnarray}
\delta f(r>\rso)&=& 
{\delta S_R\over\gamma}
\e^{i\omega\int_R^r{\d r\over \upsilon}}
\nonumber\\ 
&\times&
\left[
D_k + (1-\M^2){\upsilon\over i\omega}{B_k\over 2} 
({\delta \tilde f^-}{\delta \tilde g_0}-{\delta \tilde f_0}{\delta \tilde g^-})
\right]
\nonumber\\ 
&-&
 {\delta S_R\over2\gamma A_R} 
\bigg\{ 
\delta f^- \!\!
\int_{\rso}^r \!\!
\e^{i\omega\int_R^r{1\over 1-\M^2}{\d r\over \upsilon}}
\nonumber \\
&\times&
{\p\over\p r}
\left[
(1-\M^2){\upsilon\over i\omega} (A_k{\delta \tilde f_0}+B_k {\delta \tilde g_0})
\right]
\!
\d r
\nonumber\\
&-&
\delta f_0 \!\!
\int_{\infty}^r \!\!
\e^{i\omega\int_R^r{1\over 1-\M^2}{\d r\over \upsilon}}
\nonumber \\
&\times&
{\p\over\p r}
\left[
(1-\M^2){\upsilon\over i\omega}(A_k{\delta \tilde f^-}+B_k {\delta \tilde g^-})
\right]  \!
\d r \!
\bigg\}
\nonumber \\
\end{eqnarray}
Thus
\begin{eqnarray}
A_{k+1}&=&
-{\p\over\p r}\left\lbrack
(1-\M^2){\upsilon\over i\omega}A_k
\right\rbrack
-
{B_k\over c^2}\left(1-{\omega_L^2\over\omega^2}\right),\\
B_{k+1}&=&
-{\p\over\p r}\left\lbrack
(1-\M^2){\upsilon\over i\omega}B_k
\right\rbrack
-
\upsilon^2A_k,\\
D_{k+1}&=&D_k
+(1-\M^2)
{\upsilon\over i\omega}B_k 
\end{eqnarray}
In consequence,
\begin{eqnarray}
&&A_2= \nonumber \\ &&
{\p\over\p r}\left\lbrack
(1-\M^2)
\left(1-{\omega_{L}^2\over\omega^{2}}\right)\right\rbrack
+
{1\over c^2}\left(1-{\omega_L^2\over\omega^2}\right){\p\over\p r}(c^{2}-\upsilon^2) \nonumber \\&&\propto r^{-2},
\\
&&B_2=
{\p\over\p r}\left\lbrack
(1-\M^2){\upsilon\over i\omega}{\p\over\p r}(c^{2}-\upsilon^2)
\right\rbrack
+
{i\omega \upsilon}
\left(1-{\omega_{L}^2\over\omega^{2}}\right)\nonumber \\ && \propto r^{-2}, \\
&&D_2=c^2-\upsilon^2
-(1-\M^2)
{\upsilon\over i\omega}{\p\over\p r}(c^{2}-\upsilon^2) 
\end{eqnarray}
Noting that $D_k(\rso)=0$, the limit of this solution at the sonic point is
\begin{eqnarray}
\delta f(\rso)&=& 
 - {\delta S_R\over2\gamma A_R}
\delta f_0(\rso) \nonumber \\ &&
\times
\int_{\infty}^{\rso}
\e^{i\omega\int_R^r{1+\M^2\over 1-\M^2}{\d r\over \upsilon}}
(A_k \delta f^-+B_k \delta g^-)
\d r. \nonumber \\
\end{eqnarray}
The energy density in the supersonic region is defined by an equation similar to the subsonic region, using a reference radius $R'<\rso$, the singular function $\delta f_{\rm sup}$ defined for $r<\rso$ and choosing the boundaries of the integral to ensure the regularity and the continuity across the sonic point:
\begin{eqnarray}
&&\delta f(r<\rso)= - {\delta S_{R'}\over2\gamma A_R}\e^{i\omega\int_{R'}^{R}{\d r\over \upsilon}}
\delta f_0(r) \nonumber \\ && 
\times
\int_{\infty}^{\rso}
\e^{i\omega\int_R^r{1+\M^2\over 1-\M^2}{\d r\over \upsilon}}
(A_k \delta f^-+B_k \delta g^-)
\d r -{\delta S_{R'}\over2\gamma A_R}
\nonumber\\
&\times&
\bigg\{ \delta f_{\rm sup}
\int_{\rso}^r
\e^{i\omega\int_{R'}^r{1+\M^2\over 1-\M^2}{\d r\over \upsilon}}
\delta f_0
\left({\p\over\p r}{1\over\M^2}+
{i\omega\over \upsilon \M^2}\right)
\d r \nonumber \\ &&
-\delta f_0
\int_{\rso}^r \e^{i\omega\int_{R'}^r{1+\M^2\over 1-\M^2}{\d r\over \upsilon}}
\delta f_{\rm sup}
\left({\p\over\p r}{1\over\M^2}+{i\omega\over \upsilon\M^2}\right)
\d r
\bigg\}, \nonumber \\
\end{eqnarray}
The pressure perturbation is deduced from equation (\ref{defgp}):
\begin{eqnarray}
&&\delta p(r<\rso)=
-{\delta S_{R'}\over2\gamma A_R}\e^{i\omega\int_{R'}^{R}{\d r\over \upsilon}}
\delta p_0(r) \nonumber \\ &&
\int_{\infty}^{\rso}
\e^{i\omega\int_R^r{1+\M^2\over 1-\M^2}{\d r\over \upsilon}}
(A_k \delta f^-+B_k \delta g^-)
\d r - {\delta S_{R'}\over2\gamma A_R}
\nonumber\\
&& \times \bigg\{ \delta p_{\rm sup}
\int_{\rso}^r
\e^{i\omega\int_{R'}^r{1+\M^2\over 1-\M^2}{\d r\over \upsilon}}
\delta f_0
\left({\p\over\p r}{1\over\M^2}+{i\omega\over \upsilon\M^2}\right)
\d r \nonumber \\ &&
-\delta p_0
\int_{\rso}^r \e^{i\omega\int_{R'}^r{1+\M^2\over 1-\M^2}{\d r\over \upsilon}}
\delta f_{\rm sup}
\left({\p\over\p r}{1\over\M^2}+{i\omega\over \upsilon\M^2}\right)
\d r
\bigg\},\nonumber \\
\label{pressure_entropy}
\end{eqnarray}
where $\delta p_0$ and $\delta p_{\rm sup}$ are pressure perturbations corresponding to the homogeneous solution $\delta f_0$ and $\delta f_\mathrm{sup}$, respectively. Note that when $r\to0$, $\M\propto r^{-1/4}$, $c\propto r^{-1/4}$ and $\upsilon\propto r^{-1/2}$ for $\gamma=4/3$,
\begin{eqnarray}
A_{1} &\equiv& - {i\omega\over \upsilon}
\left(1-{\omega_{l}^2\over\omega^{2}}\right)\propto r^{-{5\over2}},\\
B_{1}&\equiv& -{\p\over\p r}(c^{2}-\upsilon^2)\propto r^{-2},\\
D_{1}&\equiv& c^2-\upsilon^2\propto r^{-1}.
\end{eqnarray}

\subsection{Continuity of the derivative of \texorpdfstring{$\delta f$}{Lg} at the sonic point}

Continuity of the derivative of $\delta f$ at the sonic point can be established in the following way. The function $\delta f$ for advected vorticity perturbations below and above the sonic point are
\begin{eqnarray}
{\delta f}(r<\rso)&=&
-{i\delta K_{R'}\over2\omega A_R} 
\nonumber \\ 
&\times&
\bigg\{ {\delta f_{\rm sup}}
\int_{\rso}^r{{\delta f_0}\over r^2\upsilon}
\e^{i\omega\int_{R'}^r{1+\M^2\over 1-\M^2}
{\d r\over \upsilon}}\d r 
\nonumber \\ 
&-&
{\delta f_0}\int_{\rso}^r{{\delta f_{\rm sup}}\over r^2\upsilon}
\e^{i\omega\int_{R'}^r{1+\M^2\over 1-\M^2}
{\d r\over \upsilon}}\d r \nonumber \\  
&-&
{\delta f_0}\e^{i\omega\int_{R'}^R{\d r\over \upsilon}}
\int_{\infty}^{\rso}{{\delta f^-}\over r^2\upsilon}
\e^{i\omega\int_{R}^r{1+\M^2\over 1-\M^2}
{\d r\over \upsilon}}\d r
\bigg\}. 
\nonumber \\ 
\\  
\delta f(r>\rso)&=&-{i\delta K_R\over2\omega A_R} \nonumber \\ 
&\times&
\bigg\{ \delta f^-
\int_{\rso}^r
\e^{i\omega\int_R^r{1+\M^2\over 1-\M^2}{\d r\over \upsilon}}
{\delta f_0\over r^2\upsilon}
\d r 
\nonumber \\ 
&-&
\delta f_0
\int_{\infty}^r
\e^{i\omega\int_R^r{1+\M^2\over 1-\M^2}{\d r\over \upsilon}}
{\delta f^-\over r^2\upsilon}
\d r 
\bigg\},
\end{eqnarray}
The derivatives of these functions are 
\begin{eqnarray}
{\partial \delta f\over\partial r}(r<\rso)&=&
-{i\delta K_{R'}\over2\omega A_R}
\nonumber \\ 
&\times&
\bigg\{ {\partial \delta f_{\rm sup}\over\partial r}
\int_{\rso}^r{{\delta f_0}\over r^2\upsilon}
\e^{i\omega\int_{R'}^r{1+\M^2\over 1-\M^2}{\d r\over \upsilon}}\d r
\nonumber \\ 
&-&
{\partial \delta f_{0}\over\partial r}
\int_{\rso}^r{{\delta f_{\rm sup}}\over r^2\upsilon}
\e^{i\omega\int_{R'}^r{1+\M^2\over 1-\M^2}
{\d r\over \upsilon}}\d r 
\nonumber \\ 
&-&
{\partial \delta f_{0}\over\partial r}\e^{i\omega\int_{R'}^R
{\d r\over \upsilon}}
\int_{\infty}^{\rso}{{\delta f^-}\over r^2\upsilon}
\e^{i\omega\int_{R}^r{1+\M^2\over 1-\M^2}{\d r\over \upsilon}}
\d r\bigg\}.\nonumber 
\\
\\
{\partial \delta f\over\partial r}(r>\rso)&=&
-{i\delta K_R\over2\omega A_R} \nonumber \\ 
&\times&
\bigg\{ 
{\partial \delta f^-\over\partial r}
\int_{\rso}^r
\e^{i\omega\int_R^r{1+\M^2\over 1-\M^2}{\d r\over \upsilon}}
{\delta f_0\over r^2\upsilon}
\d r
\nonumber \\ 
&-&
{\partial \delta f_0\over\partial r}
\int_{\infty}^r
\e^{i\omega\int_R^r{1+\M^2\over 1-\M^2}{\d r\over \upsilon}}
{\delta f^-\over r^2\upsilon}
\d r 
\bigg\}.
\end{eqnarray}
We note that the Wronskien of $(\delta f_0,\delta f_{\rm sup})$ equals that of $(\delta f_0,\delta f^-)$ except for the boundary $R$ or $R'$.
\begin{eqnarray}
{\delta f_0}{\p {f^-}\over\p r}-{\delta f^-}{\p{f_0}\over\p r}
=
-{2i\omega A_R\upsilon\over1-\M^2}\e^{-2i\omega\int_R^r{\M^2\over1-\M^2}{\d r\over 
\upsilon}},\\
{\delta f_0}{\p {\delta f_{\rm sup}}\over\p r}-
{\delta f_{\rm sup}}{\p{\delta f_0}\over\p r}
=
-{2i\omega A_R\upsilon\over1-\M^2}\e^{-2i\omega\int_{R'}^r{\M^2\over1-\M^2}{\d r\over 
\upsilon}}.
\end{eqnarray}
Thus
\begin{eqnarray}
{\p {\delta f^-}\over\p r}
=
{\delta f^-\over \delta f_0}{\p{\delta f_0}\over\p r}
-{2i\omega A_R\upsilon\over1-\M^2}{1\over \delta f_0}\e^{-2i\omega\int_R^r{\M^2\over1-\M^2}{\d r\over 
\upsilon}},\\
{\p {\delta f_{\rm sup}}\over\p r}
=
{\delta f_{\rm sup}\over \delta f_0}{\p{\delta f_0}\over\p r}
-{2i\omega A_R\upsilon\over1-\M^2}{1\over \delta f_0}\e^{-2i\omega\int_{R'}^r{\M^2\over1-\M^2}{\d r\over 
\upsilon}}.
\end{eqnarray}
Using the Wronskien relation, the derivative is rewritten as:
\begin{eqnarray}
&&
{\partial \delta f\over\partial r}(r<\rso)=-{i\delta K_{R'}\over2\omega A_R}
\nonumber \\ 
&&
\times
\bigg\{
\bigg({\delta f_{\rm sup}\over \delta f_0}{\p{\delta f_0}\over\p r}
-{2i\omega A_R\upsilon\over1-\M^2}{1\over \delta f_0}\e^{-2i\omega\int_{R'}^r{\M^2\over1-\M^2}{\d r\over 
\upsilon}}
\bigg)
\nonumber \\ 
&&
\times
\int_{\rso}^r{{\delta f_0}\over r^2\upsilon}
\e^{i\omega\int_{R'}^r{1+\M^2\over 1-\M^2}{\d r\over \upsilon}}\d r 
\nonumber \\  
&&
- 
{\partial \delta f_{0}\over\partial r}
\int_{\rso}^r{{\delta f_{\rm sup}}\over r^2\upsilon}
\e^{i\omega\int_{R'}^r{1+\M^2\over 1-\M^2}{\d r\over \upsilon}}\d r
\nonumber \\ 
&& 
-{\partial \delta f_{0}\over\partial r}\e^{i\omega\int_{R'}^R{\d r\over \upsilon}}
\int_{\infty}^{\rso}{{\delta f^-}\over r^2\upsilon}
\e^{i\omega\int_R^r{1+\M^2\over 1-\M^2}{\d r\over \upsilon}}\d r
\bigg\}.
\end{eqnarray}
\begin{eqnarray}
&&
{\partial \delta f\over\partial r}(r>\rso)=
-{i\delta K_R\over2\omega A_R} 
\nonumber \\ 
&&
\times
\bigg\{
\left({\delta f^-\over \delta f_0}{\p{\delta f_0}\over\p r}
-{2i\omega A_R\upsilon\over1-\M^2}{1\over \delta f_0}\e^{-2i\omega\int_R^r{\M^2\over1-\M^2}{\d r\over 
\upsilon}}\right)
\nonumber \\ 
&&
\int_{\rso}^r
\e^{i\omega\int_R^r{1+\M^2\over 1-\M^2}{\d r\over \upsilon}}
{\delta f_0\over r^2\upsilon}
\d r  
\nonumber \\  
&&
-{\partial \delta f_0\over\partial r}
\int_{\infty}^r
\e^{i\omega\int_R^r{1+\M^2\over 1-\M^2}{\d r\over \upsilon}}
{\delta f^-\over r^2\upsilon}
\d r
\bigg\}.
\end{eqnarray}
The limit of the derivative at the sonic point
\begin{eqnarray}
{\partial \delta f\over\partial r}(\rso^-)&=&{i\delta K_{R'}\over2\omega A_R} 
\nonumber \\ 
&\times&
\bigg\{ 
{\partial \delta f_{0}\over\partial r}(\rso)\e^{i\omega\int_{R'}^R{\d r\over \upsilon}}
\int_{\infty}^{\rso}{{\delta f^-}\over r^2\upsilon}
\e^{i\omega\int_R^r{1+\M^2\over 1-\M^2}{\d r\over \upsilon}}\d r 
\nonumber \\  
&+&
{2i\omega A_R\upsilon\over \delta f_0}{\rm lim}_{r\to\rso^-}{\e^{-2i\omega\int_{R'}^r{\M^2\over1-\M^2}{\d r\over \upsilon}}
\over 1-\M^2}
\nonumber \\ 
&\times&
\int_{\rso}^r{{\delta f_0}\e^{i\omega\int_{R'}^{r'}{\d r\over \upsilon}}\over r^2\upsilon}
\e^{2i\omega\int_{R'}^{r'}{\M^2\over 1-\M^2}{\d r\over \upsilon}}\d r'
\bigg\}.
\end{eqnarray}
\begin{eqnarray}
{\partial \delta f\over\partial r}(\rso^+)&=&{i\delta K_R\over2\omega A_R} 
\nonumber \\ 
&\times&
\bigg\{
{\partial \delta f_0\over\partial r}(\rso)
\int_{\infty}^{\rso} \, 
{\delta f^-\over r^2\upsilon} \,
\e^{i\omega\int_R^r{1+\M^2\over 1-\M^2}{\d r\over \upsilon}}
\d r \nonumber \\ 
&+&
{2i\omega A_R\upsilon\over f_0} 
{\rm lim}_{r\to\rso^+}
{\e^{-2i\omega\int_R^r{\M^2\over1-\M^2}{\d r\over \upsilon}}\over 1-\M^2}
\nonumber \\ 
&\times&
\int_{\rso}^r
{\delta f_0\e^{i\omega\int_{R}^{r'}{\d r\over \upsilon}}\over r^2\upsilon}
\e^{2i\omega\int_{R}^{r'}{\M^2\over 1-\M^2}{\d r\over \upsilon}}\d r'
\bigg\}, 
\nonumber \\ 
\end{eqnarray}
or 
\begin{eqnarray}
{\partial \delta f\over\partial r}(\rso^-)&=&
{i\delta K_{R}\over2\omega A_R}
\nonumber \\ &&
\times
\bigg\{
{\partial \delta f_{0}\over\partial r}(\rso)
\int_{\infty}^{\rso}{{\delta f^-}\over r^2\upsilon}
\e^{i\omega\int_R^r{1+\M^2\over 1-\M^2}{\d r\over \upsilon}}\d r \nonumber \\  &&
+
{2i\omega A_R\upsilon\over \delta  f_0}{\rm lim}_{r\to\rso^-}
{1\over 1-\M^2}
\nonumber \\ &&
\times
\int_{\rso}^r{{\delta f_0}\e^{i\omega\int_{R}^{r'}{\d r\over \upsilon}}\over r^2\upsilon}
\e^{2i\omega\int_{r}^{r'}{\M^2\over 1-\M^2}{\d r\over \upsilon}}\d r'
\bigg\}. \nonumber \\
\end{eqnarray}
\begin{eqnarray}
{\partial \delta f\over\partial r}(\rso^+)&=&
{i\delta K_R\over2\omega A_R}
\nonumber \\ &&
\times
\bigg\{
{\partial \delta f_0\over\partial r}(\rso)
\int_{\infty}^{\rso}{\delta f^-\over r^2\upsilon}
\e^{i\omega\int_R^r{1+\M^2\over 1-\M^2}{\d r\over \upsilon}}
\d r \nonumber \\ &&
+{2i\omega A_R\upsilon\over \delta f_0}{\rm lim}_{r\to\rso^+}
{1\over 1-\M^2}
\nonumber \\ &&
\times
\int_{\rso}^r
{\delta f_0\e^{i\omega\int_{R}^{r'}{\d r\over \upsilon}}\over r^2\upsilon}
\e^{2i\omega\int_{r}^{r'}{\M^2\over 1-\M^2}{\d r\over \upsilon}}\d r'
\bigg\}, \nonumber \\
\end{eqnarray}
We note that the right and left limit of the last term in the braces are equal:
\begin{eqnarray}
&&{\rm lim}_{x\to0^-}{1\over x}\int_0^x \e^{i\alpha \log {x'\over x}}\d x' \nonumber\\
&&=
{1\over x^{1+i\alpha}} \left\lbrack{1\over i\alpha+1}\left(x'\right)^{i\alpha+1}\right\rbrack_0^x,\nonumber\\
&&=
{1\over i\alpha+1},\nonumber\\
&&=
{\rm lim}_{x\to0^+}{1\over x}\int_0^x \e^{i\alpha \log {x'\over x}}\d x'
\end{eqnarray}
Thus the derivative of $\delta f$ is continuous across the sonic point. Using a similar procedure, we can prove the continuity of $\delta f$ for advected entropy waves.

\section{Calculation of vorticity parameters}
\label{sec:deltaK}

We first establish how components $\delta \upsilon_\bot $ and $\delta \upsilon_\mathrm{rot}$ of veloccitiy decomposition (\ref{eq:vdecompose1}) are related to the $\theta$ and $\phi$ components of velocity. For that, we substitute (\ref{eq:nabla_t1}) into equation (\ref{eq:vdecompose1}):
\begin{eqnarray}
\delta {\bm{\upsilon}} &=& \delta \upsilon_r  Y_{\ell m} {\bm{\hat r}} 
+ \delta \upsilon_\bot \left( \hat {\bm \theta} \frac{\p Y_{\ell m}}{\p \theta} + 
\hat {\bm \phi} \frac{1}{\sin\theta} \frac{\p Y_{\ell m}}{\p \phi} \right) L^{-1} \nonumber \\ \nonumber
&-& \delta \upsilon_\mathrm{rot}\, {\bm{\hat r}} \times \left( \hat {\bm \theta} \frac{\p Y_{\ell m}}{\p \theta} + 
\hat {\bm \phi} \frac{1}{\sin\theta} \frac{\p Y_{\ell m}}{\p \phi} \right)  L^{-1} \\ \nonumber
&=& \delta \upsilon_r Y_{\ell m} {\bm{\hat r}} \nonumber \\
&+&  L^{-1} \left[ \delta \upsilon_\bot \frac{\p Y_{\ell m}}{\p \theta} 
+ \delta \upsilon_\mathrm{rot} \frac{1}{\sin\theta} \frac{\p Y_{\ell m}}{\p \phi} \right] \hat {\bm \theta}  \nonumber \\
&+&  L^{-1} \left[ \delta \upsilon_\bot \frac{1}{\sin\theta} \frac{\p Y_{\ell m}}{\p \phi} 
- \delta \upsilon_\mathrm{rot} \frac{\p Y_{\ell m}}{\p \theta} \right] \hat {\bm \phi} ,
\end{eqnarray}
where we used relations ${\bm{\hat r}} \times \hat {\bm \theta} = \hat {\bm \phi}$ and ${\bm{\hat r}} \times \hat {\bm \phi} =  - \hat {\bm \theta}$. Thus
\begin{eqnarray}
\label{eq:vthetaformula}
\delta \upsilon_\theta &=& L^{-1} \left[
\delta \upsilon_\bot \frac{\p Y_{\ell m}}{\p \theta} + \delta \upsilon_\mathrm{rot} \frac{1}{\sin\theta} \frac{\p Y_{\ell m}}{\p \phi} \right], \\ 
\label{eq:vphiformula}
\delta \upsilon_\phi &=& L^{-1} \left[
\delta \upsilon_\bot \frac{1}{\sin\theta} \frac{\p Y_{\ell m}}{\p \phi} - \delta \upsilon_\mathrm{rot} \frac{\p Y_{\ell m}}{\p \theta} \right].
\end{eqnarray}
A system of differential equations for $\delta \upsilon_\theta$ and $\delta \upsilon_\phi$ can be obtained by linearizing equation~(\ref{Euler}):
\begin{eqnarray}
\label{eq:vtheta}
{\delta \upsilon_\theta\over \upsilon}
& = &
{{\omega}_\varphi\over i\omega}+
{1\over i\omega r\upsilon}
{\p\over\p\theta}\delta f-{c^2\over i\omega r\upsilon}{\p\over\p\theta}{\delta 
S\over\gamma}\eiwv,\\ 
\label{eq:vphi}
{\delta \upsilon_\varphi\over \upsilon}
& = &
-{{\omega}_\theta\over i\omega}+
{1\over i\omega r\upsilon\sin\theta}\left\lbrack
{\p\over\p\varphi}\delta f-c^2{\p\over\p\varphi}{\delta S\over\gamma}
\eiwv\right\rbrack\! , \hspace{0.7cm}
\end{eqnarray}
where $\omega_\theta$ and $\omega_\varphi$ are the $\theta$ and $\varphi$ components of the vorticity perturbation, which can be obtained from the linearized vorticity equation (\ref{eq:vort}) \citep{kovalenko:98,foglizzo:01}:
\begin{eqnarray}
\omega_\theta&=&{1\over r\upsilon}  \left\lbrack R\upsilon_R(\omega_\theta)_R -
{c^2-c_R^2\over\sin\theta}{\p\over\p\varphi}{\delta 
S_{R}\over\gamma}\right\rbrack\! \eiwv,\label{wtet}\\
\omega_\varphi&=&{1\over r\upsilon} \left\lbrack R\upsilon_R(\omega_\varphi)_R +
(c^2-c_R^2){\p\over\p\theta}
{\delta S_{R}\over\gamma}\right\rbrack\! \eiwv.\qquad \label{wphi}
\end{eqnarray}
Equations (\ref{eq:vtheta}) and (\ref{eq:vphi}) can be combined into
\begin{equation}
\frac{r}{\sin\theta}\left[
\frac{\partial}{\partial \theta} \left( \sin \theta \delta \upsilon_\theta \right) +
 \frac{\partial}{\partial \phi} \delta \upsilon_\phi  \right] = 
 \frac{1}{i \omega } \left[\delta K - L^2 \delta f \right] Y_{\ell m},
\end{equation}
where $L^2\equiv l(l+1)$. Using formulas (\ref{eq:vthetaformula})-(\ref{eq:vphiformula}), we can obtain 
\begin{equation}
\label{eq:vt}
\frac{r}{\sin\theta}\left[
\frac{\partial}{\partial \theta} \left( \sin \theta \delta \upsilon_\theta \right) +
 \frac{\partial}{\partial \phi} \delta \upsilon_\phi  \right] = - L r \delta \upsilon_\bot Y_{\ell m}.
\end{equation}
Combining the last two equations, we obtain an expression for $\delta \upsilon_\bot$:
\begin{equation}
\label{eq:dv_bot}
\delta \upsilon_\bot = \frac{L}{i \omega r} \left( \delta f - \frac{\delta K}{L^2} \right).
\end{equation}
Next, we decompose the vorticity vector into vector spherical harmonics:
\begin{equation}
\label{eq:vorticity1}
\delta {\boldsymbol{\omega}} = \delta \omega_r Y_{\ell m} {\bm{\hat r}} 
+ \delta \omega_\bot L^{-1} \boldsymbol{\hat \nabla}_\bot Y_{\ell m} 
- \delta \omega_\mathrm{rot} \, L^{-1} {\bm{\hat r}} \times \boldsymbol{\hat \nabla}_\bot Y_{\ell m} 
\end{equation}
The vorticity perturbation can be calculated as \citep[e.g.,][]{lai:00}
\begin{eqnarray}
\label{eq:vorticity2}
\delta {\boldsymbol{\omega}} &=& \nabla \times \bm{\delta \upsilon} = 
\frac{L}{r} \delta \upsilon_\mathrm{rot} Y_{\ell m}  \bm{\hat r} 
\nonumber \\ 
&+&
\frac{1}{r} \p_r \left( r \delta \upsilon_\mathrm{rot} \right) L^{-1} \boldsymbol{\hat \nabla}_\bot Y_{\ell m} 
\nonumber \\ 
&-& \frac{L \delta \upsilon_r - \p_r (r \delta \upsilon_\bot)}{r} L^{-1} {\bm{\hat r}} \times \boldsymbol{\hat \nabla}_\bot Y_{\ell m}. 
\end{eqnarray} 
We now apply this formula to calculate the radial component of $\nabla \times \delta {\boldsymbol{\omega}}$:
\begin{equation}
\left( \nabla \times \bm{\delta {\boldsymbol{\omega }}} \right)_r= 
\frac{L}{r} \delta \omega_\mathrm{rot} Y_{\ell m}
\end{equation} 
Thus, in the absence of entropy perturbations,
\begin{equation}
\delta K = r^2 \upsilon_r \left(\nabla \times \delta { \boldsymbol{\omega}} \right)_r = 
L r \upsilon_r \delta \omega_\mathrm{rot} Y_{\ell m},
\end{equation}
which is valid in linear order in the perturbation magnitude. The component $\delta \omega_\mathrm{rot}$ can be obtained by comparing equation (\ref{eq:vorticity1}) and (\ref{eq:vorticity2}):
\begin{equation}
\delta \omega_\mathrm{rot} = \frac{L \delta \upsilon_r - \p_r (r \delta \upsilon_\bot)}{r},
\end{equation}
which leads to the following expression for $\delta K$
\begin{equation}
\delta K = L \upsilon \left[L \delta \upsilon_r - \p_r (r \delta \upsilon_\bot) \right] Y_{\ell m}.
\end{equation}

\section{Relation between the dimensionless entropy and the entropy per nucleon}
\label{sec:entropy}

In this section, we derive a relation between the dimensionless entropy that we use and the entropy per nucleon that is usually used in the literature on CCSNe. We use the thermodynamic relation 
\begin{equation}
\label{eq:ds}
ds = \gamma c_v \left( \frac{dp}{\gamma p} - \frac{d\rho}{\rho} \right),
\end{equation}
where $ds$ is the specific entropy and $c_v$ is the specific heat at constant volume. Using the relation 
\begin{equation}
\label{eq:rconstant}
c_v = \frac{1}{\mu}\frac{R}{\gamma-1},
\end{equation}
where $R$ is the universal gas constant and $\mu$ is the molar mass, equation (\ref{eq:ds}) is re-written as
\begin{equation}
ds = \frac{\gamma}{\gamma-1} \frac{R}{\mu} \left( \frac{dp}{\gamma p} - \frac{d\rho}{\rho} \right).
\end{equation}
The entropy is made dimensionless by setting $R/\mu=1$ without loss of generality: 
\begin{equation}
dS = \frac{\gamma}{\gamma-1} \left( \frac{dp}{\gamma p} - \frac{d\rho}{\rho} \right),
\end{equation}
where we used $dS$ to denote the dimensionless entropy. The entropy per nucleon, which we denote as $ds_\mathrm{b}$, is related to the specific entropy $ds$ via
\begin{equation}
ds = \frac{ds_\mathrm{b}}{m_\mathrm{b}}.
\end{equation}
Thus, 
\begin{equation}
dS = \frac{ds_\mathrm{b} \mu}{R m_\mathrm{b}}, 
\end{equation}
Since 
\begin{equation}
\frac{\mu}{ Rm_\mathrm{b}}=\frac{N_\mathrm{A}}{R} = \frac{1}{k_\mathrm{b}},
\end{equation}
where $k_\mathrm{b}$ is the Boltzmann constant, we obtain
\begin{equation}
\label{eq:entropy}
dS = \frac{ds_\mathrm{b} }{k_\mathrm{b}}, 
\end{equation}
which gives us a relation between the dimensionless entropy and the entropy per nucleon.

\section{Decomposition of hydrodynamic perturbations into physical components}
\label{sec:decomposition}

For uniform inviscid mean flow, the acoustic, entropy, and vorticity perturbations evolve independently from each other in the linear approximation \citep{kovasznay:53}. However, this is no longer the case for non-uniform background flow. Nevertheless, we can separate the vorticity waves using incompressibility condition, while the in-going and out-going acoustic waves can be separated using the WKB approximation \citep{foglizzo:07}. In this approach, we decompose perturbations at a given point assuming the perturbations are allowed to evolve in a uniform flow at the same point:  
\begin{eqnarray}
\delta f &=& \delta f^+ + \delta f^- + \delta f^S + \delta f^K, \\
\delta g &=& \delta g^+ + \delta g^- + \delta g^S + \delta g^K,
\end{eqnarray}
where $\delta f^+$ and $\delta f^-$ are the contributions of ingoing and outgoing acoustic waves, while $\delta f^S$ and $\delta f^K$ correspond to $\delta S$ and $\delta K$, which is given as\footnote{\citet{foglizzo:07} uses function $h$, which related to our function $\delta g$ through the equation $h\equiv \delta g-\delta S$.} 
\begin{eqnarray}
\delta f^K   &\equiv& \frac{{\cal M}^2 (1-\mu^2)}{1-{\cal M}^2 \mu^2} \frac{\delta K}{L^2}, \\
\delta g^K   &\equiv& \frac{\delta f^K}{\upsilon^2} + \delta S, \\
\delta f^S   &\equiv& \frac{c^2(1-{\cal M}^2)}{1-\mu^2 {\cal M}^2} \frac{\delta S}{\gamma}, \\ 
\delta g^S   &\equiv& \frac{\mu^2}{c^2} \delta f^S, \\
\delta f^\pm &\equiv& \frac{1}{2}\delta f \pm \frac{{\cal M} c^2}{2 \mu} \delta g -
 \frac{1 \pm \mu{\cal M}}{2}\left(\delta f^S\pm \frac{\delta f^K}{\mu{\cal M}}\right),
\end{eqnarray}
where
\begin{equation}
\mu^2 \equiv 1 - \frac{L^2 c^2}{\omega^2 r^2}\left(1-{\cal M}^2\right).
\end{equation}
Note that the decomposition of acoustic waves into ingoing and outgoing waves is valid only in the WKB regime where the wavelength of the perturbations is much smaller than the characteristic scale of the background flow. 
The corresponding values of the perturbations of velocity, density, and pressure are obtained from formulas (\ref{deffvr})-(\ref{defgp}). For vorticity waves, $\delta K\ne 0$ and $\delta S=0$, which leads to 
\begin{eqnarray}
\label{eq:vr_vor}
\frac{\delta \upsilon_r}{\upsilon}    &=& \frac{1}{\upsilon^2} \frac{{\cal M}^2 (1-\mu^2)}{1-\mu^2 {\cal M}^2} \frac{\delta K}{L^2}, \\
\frac{\delta \upsilon_\bot}{\upsilon} &=& \frac{1}{i \omega r \upsilon} \frac{{\cal M}^2-1}{1-\mu^2 {\cal M}^2 } \frac{\delta K}{L}
\label{eq:vt_vor}
\end{eqnarray}
The density and pressure change are zero for vorticity waves in a uniform background flow. For entropy waves, we linearly superpose two solutions with $\delta K = c_R^2 L^2 \delta S / \gamma $ and $\delta S \ne 0$, where $R$ is the initial radius of entropy perturbations. The velocity of the vorticity waves generated by advected entropy perturbations are 
\begin{eqnarray}
\label{eq:vr_ent}
\frac{\delta \upsilon_r}{\upsilon}    &=& \frac{1-\mu^2}{1-\mu^2 {\cal M}^2} \left(\frac{c_R^2}{c^2}-1 \right) \frac{\delta S}{\gamma}, \\
\frac{\delta \upsilon_\bot}{\upsilon} &=& \frac{i L}{\omega r \upsilon} \frac{1-{\cal M}^2}{1-\mu^2 {\cal M}^2} \left(c_R^2-c^2 \right) \frac{\delta S}{\gamma}.
\label{eq:vt_ent}
\end{eqnarray}
The associated pressure perturbations is zero because entropy perturbations do not produce pressure variation in a uniform background flow. The associated density perturbations can be obtained from the thermodynamic relation (\ref{eq:ds}).

In the limit $r\rightarrow 0$, $\upsilon \propto r^{-1/2}$ and $c \propto r^{-1/4}$ for $\gamma=4/3$, which results in ${\cal M} \propto r^{-1/4}$ and $\mu^2 \propto r^{-3}$. For the velocity field of vorticity waves (\ref{eq:vr_vor})-(\ref{eq:vt_vor}), this results in $\delta \upsilon_r \propto r^{1/2}$ and $\delta \upsilon_\bot \propto r^2$. For the vorticity waves generated by advected entropy waves (\ref{eq:vr_ent})-(\ref{eq:vt_ent}), we obtain  $\delta \upsilon_\bot \propto r^{3/2}$ and $\delta \upsilon_r \propto \mathrm{const}$.

\bsp	
\label{lastpage}
\end{document}